\documentclass[a4paper,prl,twocolumn,superscriptaddress]{revtex4-1}

\usepackage{amssymb}
\usepackage{amsmath}
\usepackage{epsfig}
\usepackage{color}
\usepackage{graphics, graphicx}
\usepackage{bbold}
\usepackage{psfrag}
\usepackage{mathcomp}
\usepackage{subfigure}
\usepackage{verbatim}
\usepackage{color}
\usepackage{blindtext}
\usepackage{float}
\usepackage[utf8]{inputenc}
\usepackage[T1]{fontenc}
\usepackage[colorlinks,citecolor=blue]{hyperref}

\begin{document}
\title{Emergence of the Unconventional Type-II Nambu-Goldstone Modes with Topological Origin in Bose Superfluids}

\author{Jian-Song Pan}
\affiliation{Wilczek Quantum Center, School of Physics and Astronomy and T. D. Lee Institute,
Shanghai Jiao Tong University, Shanghai 200240, China}
\affiliation{International Center for Quantum Materials, School of Physics, Peking University, Beijing 100871, China}

\author{W. Vincent Liu}
\email{wvliu@pitt.edu}
\affiliation{Department of Physics and Astronomy, University of Pittsburgh, Pittsburgh, Pennsylvania 15260, USA}
\affiliation{Wilczek Quantum Center, School of Physics and Astronomy and T. D. Lee Institute, Shanghai Jiao Tong University, Shanghai 200240, China}
\affiliation{Shenzhen Institute for Quantum Science and Engineering and Department of Physics, Southern University of Science and Technology, Shenzhen 518055, China}

\author{Xiong-Jun Liu}
\email{xiongjunliu@pku.edu.cn}
\affiliation{International Center for Quantum Materials, School of Physics, Peking University, Beijing 100871, China}
\affiliation{CAS Center for Excellence in Topological Quantum Computation, University of Chinese Academy of Sciences, Beijing 100190, China}
\affiliation{Shenzhen Institute for Quantum Science and Engineering and Department of Physics, Southern University of Science and Technology, Shenzhen 518055, China}
\affiliation{Synergetic Innovation Center for Quantum Effects and Applications, Hunan Normal University, Changsha 410081, China}

\begin{abstract}
The Nambu-Goldstone (NG) modes in a nonrelativistic system can be classified into two types from their characteristic features: being of either an odd (type I) or an even (type II) power energy-momentum dispersion. Conventionally, the type-II NG modes may universally arise from spontaneous breaking of noncommutative symmetry pairs. Here, we predict a novel type of quadratically dispersed NG modes that emerges in mixed $s$ and $p$ band Bose superfluids in an optical lattice and, unlike the conventional type-II NG modes, cannot be solely interpreted with the celebrated symmetry-based argument. Instead, we show that the existence of such modes has a profound connection to the topological transition on projective complex order-parameter space. The detection scheme is also proposed. Our Letter reveals a new universal mechanism for emergence of type-II NG modes, which bridges intrinsically the Landau symmetry-breaking and topological theories. 
\end{abstract}

\maketitle
\emph{Introduction}.--
Characterizing quantum phases within a unified framework is the basic pursuit of condensed matter physics.
One of the most fundamental notions is the Landau's symmetry-breaking paradigm, which characterizes quantum phases by order parameters breaking certain symmetries under consideration~\cite{landau1937theory,ter2013collected}. The low-energy physics of symmetry-breaking phases are captured by the emergent Nambu-Goldstone (NG) modes~\cite{goldstone1962broken}. In general, the type-I NG modes with odd-power dispersion appear in a phase that breaks a continuous symmetry, while type-II NG modes with even-power dispersion appear when a pair of noncommutative symmetries are simultaneously broken~\cite{nielsen1976count,watanabe2011number,watanabe2012unified,hidaka2013counting,watanabe2014effective, leutwyler1994nonrelativistic,schafer2001kaon, takahashi2015counting, type_AB}. A notable example is that the magnons in ferromagnets have quadratic dispersion due to breaking two noncommutative rotation symmetries by spontaneous magnetization~\cite{Bloch1930Zur}.

Topological quantum phase is another fundamental notion, brought about from the discovery of integer quantum Hall effect and beyond the symmetry-breaking framework, and are classified by global topological invariants~\cite{wen2004quantum}. The pursuit of topological matter has been remarkably revived in the past over ten years due to the prediction and discovery of various fundamental types of topological phases in condensed matter physics, such as topological insulators and topological superconductors~\cite{hasan2010colloquium,qi2011topological,shen2012topological,hasan2011three,chiu2016classification,bansil2016colloquium, wen2017colloquium}, with considerable efforts having been also made in ultracold atoms~\cite{jotzu2014experimental,aidelsburger2015measuring,wu2016realization,meng2016experimental,lohse2018exploring,song2018observation,song2018observation,cooper2019topological,LU2020,wang2020realization}. For a gapped topological phase, the boundary hosts gapless excitations which have odd-power energy-momentum dispersions.

In this Letter, we predict an unconventional type of quadratically dispersed NG modes for mixed $s$ and $p$ band Bose superfluids in a two-dimensional (2D) optical lattice, and show that such type-II NG modes 
manifest a novel paradigm that intrinsically bridges the Landau symmetry-breaking and topological theories. The low-energy physics of the superfluids depend on the breaking of the global gauge symmetry, rotation, and/or time-reversal symmetry~\cite{liu2006atomic, li2016physics}.
The former two symmetries commute, hence no conventional type-II NG modes are expected. Surprisingly, we show a branch of type-II NG modes emerging at a phase-diagram boundary, across which the broken symmetries are the same but the degenerate space of ground states undergoes a topological transition after being projected onto the complex order-parameter subspace. This prediction shows a fundamentally new type-II NG mode with topological origin, and the detection scheme is also proposed.

\begin{figure*}[tbp]
\includegraphics[width=17.5cm]{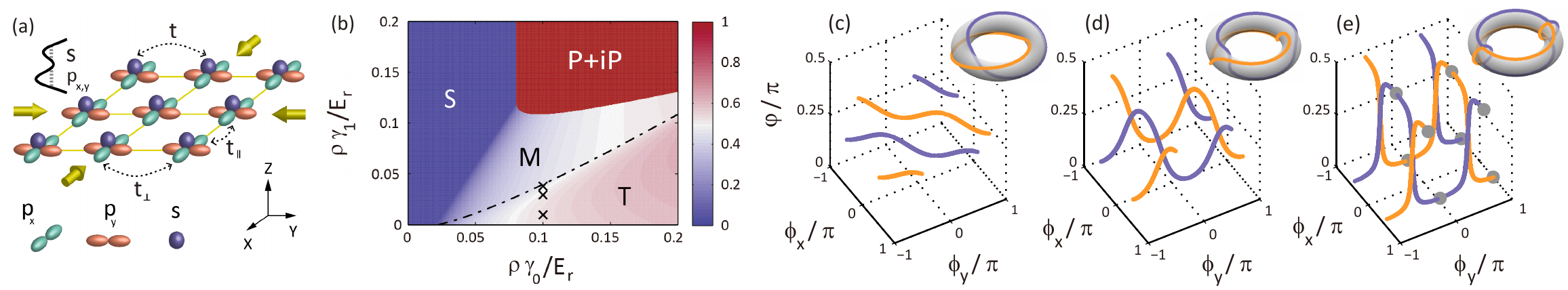}
\caption{(a) Illustration of a quasi-2D optical lattice trapped with double-well potential in $z$ direction. Only the $s$ orbital of the shallower layer and $p$ orbitals of the deeper layer are involved in realization. (b) The phase diagram on the $\gamma_0$-$\gamma_1$ plane, where the background color shows the density of the $p$ bands ($\sim\sin^2\theta$). The density ratio varies continuously between subphases $M$ and $T$, which are distinguished by different phase windings (see the main text). (c)-(e) The degenerate order-parameter trajectories generated by the $\text{SO(2)}$ symmetry at three typical parameter points in subphase $T$ [(c) $\rho\gamma_0/E_r=0.1$ and $\rho\gamma_1/E_r=0.01$, (d) $0.035$, and (e) $0.039$ with the average particle density $\rho$; see the crosses in (b)]. Different colors distinguish the twofold degeneracy generated by the time-reversal symmetry. Insets of (c)-(e): the phase winding of the order parameters are illustrated on a torus, where the rotation angles around the horizontal and vertical rotation axes represent $\phi_x$ and $\phi_y$, respectively. The gray markers in (e) denote the positions where the $p_x$ or $p_y$ components of the order parameter will vanish due to the breaking of phase winding at the $T$-$M$ boundary. Here we set $t/E_r=t_{\perp}/E_r=0.02$, $t_{\parallel}=0.2$, $\tilde{\Delta}/E_r=[4t-2(t_{\parallel}+t_{\perp})+\Delta]/E_r=0.02$, where $E_{r}=\hbar^2k_0^2/2m$ is the recoil energy with the lattice wave vector $k_0$ and atomic mass $m$.}
\label{fig:PD}
\end{figure*}

\emph{The model}.--We start with a concrete system of weakly interacting bosons on the lowest excited bands of a quasi-2D square lattice in $x-y$ plane, trapped by a double-well potential in $z$ direction. This includes nearly degenerate $p$ bands of the deeper layer and $s$ band of the shallower layer [Fig.~\ref{fig:PD}(a)], and is experimentally realizable~\cite{wirth2011evidence, olschlager2013interaction, kock2015observing}. Denote by creation operators at $\boldsymbol{j}$th site $\hat{\psi}_{z\boldsymbol{j}}^{\dagger}$ for $s$ orbital and $\hat{\psi}_{\nu=x,y\boldsymbol{j}}^{\dagger}$ for ${p}_{\nu}$ orbital with odd parity in the $\nu$ direction. The Hamiltonian $H=H_{T}+H_0$ reads
\begin{eqnarray}
H_T&=&-\sum_{\langle\boldsymbol{i},\boldsymbol{j}\rangle\nu=x,y} (t_{\boldsymbol{i}\boldsymbol{j}}\hat{\psi}_{\nu\boldsymbol{i}}^{\dagger}\hat{\psi}_{\nu\boldsymbol{j}}+t\hat{\psi}_{z\boldsymbol{i}}^{\dagger}\hat{\psi}_{z\boldsymbol{j}}),\label{eq_H0Vint}\\
H_{0}&=&\frac{1}{2}:\sum_{\boldsymbol{j}}[\gamma_{0}\hat{n}_{\boldsymbol{j}}^{(z)2}+\gamma_{1}(3\hat{n}_{\boldsymbol{j}}^{\left(z\right)}\hat{n}_{\boldsymbol{j}}^{\left(xy\right)}-\frac{1}{2}\sum_{\nu=x,y}\hat{L}_{\nu,\boldsymbol{j}}^{2})\nonumber\\
&&+\frac{3}{4}\gamma_{2}(\hat{n}_{\boldsymbol{j}}^{(xy)2}-\frac{1}{3}\hat{L}_{z,\boldsymbol{j}}^{2})+\Delta\hat{n}_{\boldsymbol{j}}^{(xy)}]:.\label{eq_VintHam}
\end{eqnarray}
Here $\hat{n}_{\boldsymbol{j}}^{\left(\nu\right)}=\hat{\psi}_{\nu\boldsymbol{j}}^{\dagger}\hat{\psi}_{\nu\boldsymbol{j}}$, $\hat{n}_{\boldsymbol{j}}^{\left(xy\right)}=\hat{n}_{\boldsymbol{j}}^{\left(x\right)}+\hat{n}_{\boldsymbol{j}}^{\left(y\right)}$, and $\hat{L}_{\varsigma=x,y,z,\boldsymbol{j}}=-i\sum_{\tau,\upsilon=x,y,z}\epsilon_{\tau\upsilon\varsigma}\hat{\psi}_{\tau\boldsymbol{j}}^{\dagger}\hat{\psi}_{\upsilon\boldsymbol{j}}$ with the Levi-Civita symbol $\epsilon_{\tau\upsilon\varsigma}$ are the local angular-momentum operators~\cite{liu2006atomic, li2016physics}. The hopping coefficient $t_{\boldsymbol{i}\boldsymbol{j}}=-t_{\parallel}$ (or $t_{\boldsymbol{i}\boldsymbol{j}}=t_{\perp}$) for the hopping of ${p}_{\nu}$ orbital along (or perpendicular to) $\nu$ direction, with $t_{\parallel}>t_{\perp}$, and $t$ is the hopping strength of $s$ orbital in $x-y$ plane. The term $H_{0}$ is written in normal order, with $\Delta$ denoting the energy difference between $s$ and $p_{x,y}$ orbitals, and $\gamma_{0,1,2}$ the interaction coefficients between different orbitals (see Supplemental Material~\cite{suplementary}). For convenience, we set $\gamma_{2}=\gamma_{0}$. The Hamiltonian $H$ possesses the global $\text{U(1)}$ gauge symmetry and time-reversal $\mathcal{T}$ symmetry~\cite{liu2006atomic, li2016physics}. Further, in the long-wave limit the system retrieves an $\text{SO(2)}$ rotation symmetry generated by $\hat{L}_z=\sum_{\boldsymbol{j}}\hat{L}_{z,\boldsymbol{j}}$.

\begin{figure*}[tbp]
\includegraphics[width=18cm]{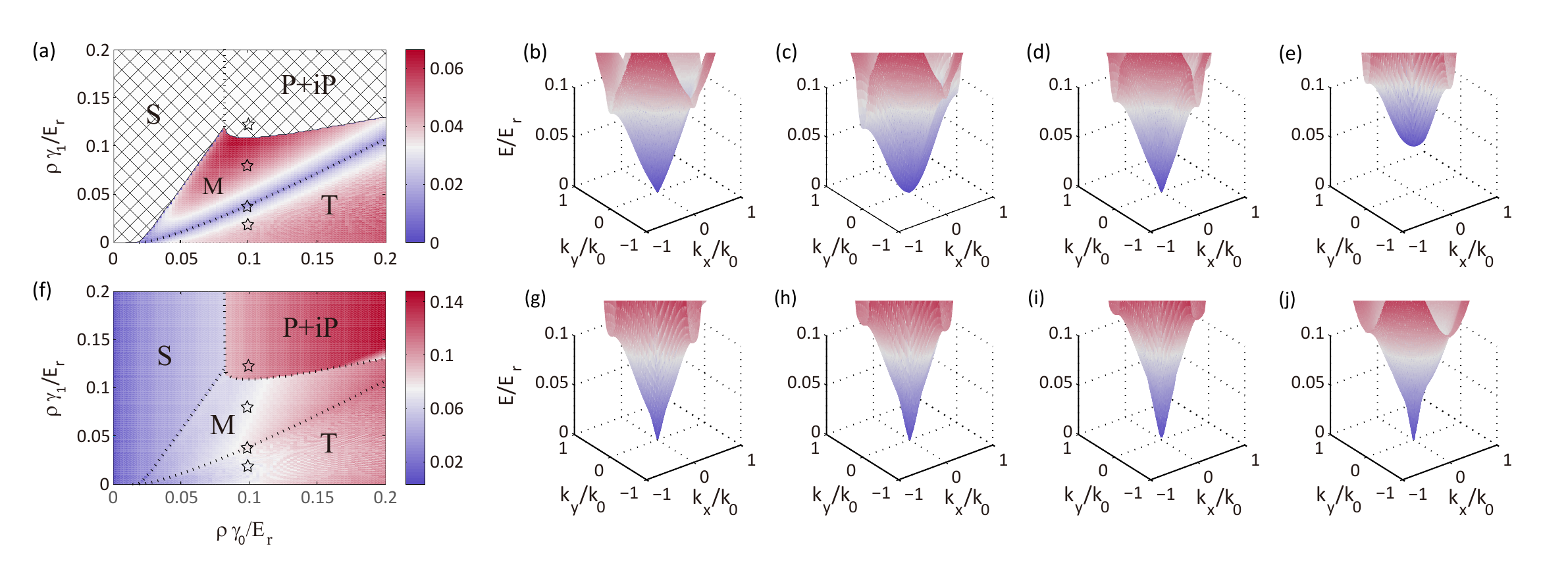}
\caption{ The NG modes associated with the $\text{SO(2)}$ (a) and $\text{U(1)}$ (f) symmetries are shown with the first-order dispersion coefficient at $k=0$ on the $\gamma_0$-$\gamma_1$ plane (the correspondence between NG modes and symmetries is identified by comparing with analytical results~\cite{suplementary}).
The shaded region in (a) corresponds to the absence of NG mode associated with the $\text{SO(2)}$ symmetry in phases $S$ and $P$+i$P$ (see the main text). Rest sub-figures: the excitation bands with NG modes associated with the $\text{SO(2)}$ (upper) and $\text{U(1)}$ (lower) symmetries at four typical parameter points $\rho\gamma_1/E_r=0.02$ (b,g), $0.04$ (c,h), $0.08$ (d,i) and $0.12$ (e,j) with fixed $\rho\gamma_0/E_r=0.1$ (see the pentagons in (a) and (f)). We find the dispersion of the NG mode associated with the $\text{SO(2)}$ symmetry becomes quadratic at the $T$-$M$ boundary, which implies a type-II NG mode emerges~\cite{suplementary}. }
\label{fig:NG MODE}
\end{figure*}

\emph{Superfluid phases}.--Without considering the interactions, the single particle energies read $\epsilon_{s}=-2t(\cos k_{x}+\cos k_{y})$ and $\epsilon_{p_{x(y)}}= t_{\parallel}\cos k_{x(y)}-t_{\perp}\cos k_{y(x)}$, which respectively have the minimums at momenta $\boldsymbol{Q}_z=(0,0)$ and ${\bold Q}_{x}[\boldsymbol{Q}_y]=(\pi,0)[(0,\pi)]$. The superfluid order parameter can then be written as
$(\langle\hat{\psi}_{z\boldsymbol{k}}\rangle, \langle\hat{\psi}_{x\boldsymbol{k}}\rangle, \langle\hat{\psi}_{y\boldsymbol{k}}\rangle)=\sqrt{N_a}(
\delta_{\boldsymbol{k},\boldsymbol{0}}\cos\theta, \delta_{\boldsymbol{k},\boldsymbol{Q}_{x}}\sin\theta\cos\varphi e^{i\phi_{x}},  \delta_{\boldsymbol{k},\boldsymbol{Q}_{y}}\sin\theta\sin\varphi e^{i\phi_{y}})$ with $N_a$ being the total atom number~\cite{li2016physics, orderparameter}. The angles $\theta, \varphi\in[0,\pi/2]$ and $\phi_x,\phi_y \in (-\pi,\pi]$ for the parameter space of the complex order are determined by minimizing the energy functional~\cite{suplementary}.

We present first the numerical results (see~\cite{suplementary} for more details). Fig.~\ref{fig:PD}(b) shows the phase diagram by minimizing the total energy functional, with the three symmetry-breaking phases denoted as S, $P+iP$, and $M\&T$.
The phase $S$ breaks only the $\text{U(1)}$ symmetry, with atoms condensed into $s$ band (i.e. $\theta=0$). In the chiral phase $P+iP$, both $\text{U(1)}$ and $\mathcal{T}$ symmetries are broken, such that atoms condense into the $p_x+ip_y$ orbital (i.e. $\theta=\pi/2$ and $\varphi=\pi/4$) with $\phi_x-\phi_y=\pm\pi/2$. In the subphases $M\&T$, the $\text{U(1)}$, $\text{SO(2)}$ and $\mathcal{T}$ symmetries are all broken.

The $M$ and $T$ subphases are however different. In subphase $M$, the ground state degeneracy generated by $\text{SO(2)}$ symmetry~\cite{suplementary} is characterized by $\varphi\in[0,\pi/2]$ with fixed $\phi_{x,y}=\pm\pi/2$ and $|\phi_{x}-\phi_y|=\pi$. If we depict $\langle\hat{\psi}_{x,\boldsymbol{Q}_{x}}\rangle$ (the same for $\langle\hat{\psi}_{y,\boldsymbol{Q}_{y}}\rangle$) in a complex plane, its degenerate-space trajectory is a straight line crossing the origin point. In subphase $T$, however, both $\phi_{x}$ and $\phi_y$ sweep over $[0,2\pi)$, and $\varphi$ varies in $[\pi/4-\Delta\varphi,\pi/4+\Delta\varphi]$ with $\Delta\varphi<\pi/4$ in the degenerate space [see Fig.~\ref{fig:PD}(c)-(e)]. The complex-plane degenerate trajectories of $p_x$ and $p_y$ components wind around the origin. Approaching the boundary between $M$ and $T$, these winding trajectories are suppressed into flat ellipses and the phase angles $\phi_x$ and $\phi_y$ tend to be ill defined at some degenerate states [see Fig.~\ref{fig:PD}(e), gray markers]. This essential difference brings about nontrivial NG modes in the $M$ and $T$ phase boundary, as we elaborate below.

\emph{Unconventional type-II NG modes}.--The gapless NG modes correspond to the spontaneously breaking symmetries~\cite{pethick2008bose}, as shown numerically in Fig.~\ref{fig:NG MODE}~\cite{suplementary}.
The type-I NG mode arises generically in the different superfluid phases due to breaking $\text{U(1)}$ symmetry [see Fig.~\ref{fig:NG MODE}(f)-(i)]. In the $M\&T$ phase, another NG mode emerges due to the breaking of $\text{SO(2)}$ symmetry [Fig.~\ref{fig:NG MODE}(a)-(e)]. Our key observation is that the NG mode corresponding to the broken $\text{SO(2)}$ symmetry softens to a type-II NG mode with quadratic dispersion at the $M$-$T$ phase boundary~[Fig.~\ref{fig:NG MODE}(a) and (c)]. Analytic perturbation analysis also confirms the emergent type-II NG mode~\cite{suplementary}.

The emergence of the type-II NG modes here is beyond the conventional interpretation~\cite{nielsen1976count,watanabe2011number,watanabe2012unified,hidaka2013counting,watanabe2014effective, leutwyler1994nonrelativistic,schafer2001kaon, hayata2018diffusive}. For the system with kinetic energies around the band bottom being regular ($\propto k^2$, with $k$ measured from band bottom), which is the case for the present $s$-$p$ superfluids, the type-II NG mode coincides with the type-B mode~\cite{NGMclassification,type_AB}. Conventionally, the numbers of type-II ($N_2$) and type-I ($N_1$) NG modes are $N_2=\text{rank}(\rho^{ab})/2$ and $N_1=N_{BS}-2N_2$. Here $N_{BS}$ is the number of generators ($\hat{Q}_a$) of the broken symmetries and matrix $\rho^{ab}\equiv\langle[i\hat{Q}_a,\hat{Q}_b]\rangle/V$ is defined by the commutators, with the system volume $V$ tending to infinity. It follows that the broken noncommutative symmetry pairs determine the type-II NG modes. However, the generators of $\text{U(1)}$ and $\text{SO(2)}$ symmetries commute, hence the emergence of the type-II NG modes is unconventional and has a profound origin: it relates to topological transition on a projective complex order-parameter space, with a general framework being developed below.

\emph{Order-parameter projection topology}.--We proceed to unveil the underlying mechanism of above prediction. We build up a generic theory {for multi-component homogeneous Bose superfluids with the low-energy Hamiltonian} $H=\int d^{3}\boldsymbol{r}[\hbar^{2}\sum_{j=1}^{N}\nabla\hat{\psi}_{j}^{\dagger}\nabla\hat{\psi}_{j}/2m+H_{0}(\hat{\Psi}^{\dagger},\hat{\Psi})]$, with
\begin{equation}\label{eq_Vint}
H_{0}=\sum_{j=1}^{N}\epsilon_{j}\hat{\psi}_{j}^{\dagger}\hat{\psi}_{j}+\sum_{i,j,k,l=1}^{N}g_{ijkl} \hat{\psi}_{i}^{\dagger}\hat{\psi}_{j}^{\dagger}\hat{\psi}_{k}\hat{\psi}_{l}.
\end{equation}
Here $m$ is the atomic mass, $\hat{\psi}_j$ denotes the field operator of the $j$ th component with single-particle energy $\epsilon_j$, and $\hat{\Psi}=(\begin{array}{ccc}
\hat{\psi}_{1} & \cdots & \hat{\psi}_{N}\end{array})^{T}$. This theory is valid for both continuous and lattice systems, and hence applicable to the model in Eqs.~(\ref{eq_H0Vint}) and (\ref{eq_VintHam})~\cite{modelconnection}. We assume the interaction coefficient $g_{ijkl}$ satisfies $g_{ijkl}=g_{lkji}$ so that $H_{0}$ is Hermitian, and multicomponent scattering is neglectable, i.e. $g_{ijkl}\approx0$ for more than two different components.
The superfluid order parameter is given by $\Psi=\langle\hat{\Psi}\rangle=(\begin{array}{ccc}
\psi_{1} & \cdots & \psi_{N}\end{array})^{T}$ with $\psi_j=\langle\hat{\psi}_j\rangle$. Here $\Psi$ is governed by the Gross-Pitaevskii (GP) equation~\cite{pethick2008bose},
\begin{equation}\label{eq_GP}
\mu\Psi=\frac{\partial H_0(\Psi^{\ast},\Psi)}{\partial\Psi^{\ast}}=(\begin{array}{ccc}
\frac{\partial H_0}{\partial\psi_{1}^{\ast}} & \cdots & \frac{\partial H_0}{\partial\psi_{N}^{\ast}}\end{array})^{T}.
\end{equation}
The chemical potential $\mu$ is consistently determined with the particle number equation $N_{a}=\sum_{j=1}^{N}\int d^{3}\boldsymbol{r}\left|\psi_{j}\right|^{2}$.

Other than $\text{U(1)}$ gauge symmetry, for the theory, we require that the Hamiltonian $H$ has another continuous symmetry characterized by the periodically parameterized symmetry group
$\mathcal{G}=\{g(\xi),\xi\in[0,2\pi)\}$, with the group elements $g(\xi)=g(\xi+2\pi)$ labeled by group parameter $\xi$. For example, the rotation symmetry around the y direction is parameterized as $\{g(\xi)=e^{i\hat{J}_{y}\xi},\xi\in[0,2\pi)\big|g(\xi)=g(\xi+2\pi)\}$, with the angular momentum operator $\hat{J}_{y}$ and rotation angle $\xi$. When this symmetry is spontaneously broken, the system picks one specific ground state with $\xi_0$ in the degenerate subspace so that $g(\xi)\Psi(\xi_0)\neq\Psi(\xi_0)$ if $\xi\neq0$, and $\xi_0$ can be any value in experimental realization.

\begin{figure}[tbp]
\includegraphics[width=8.5cm]{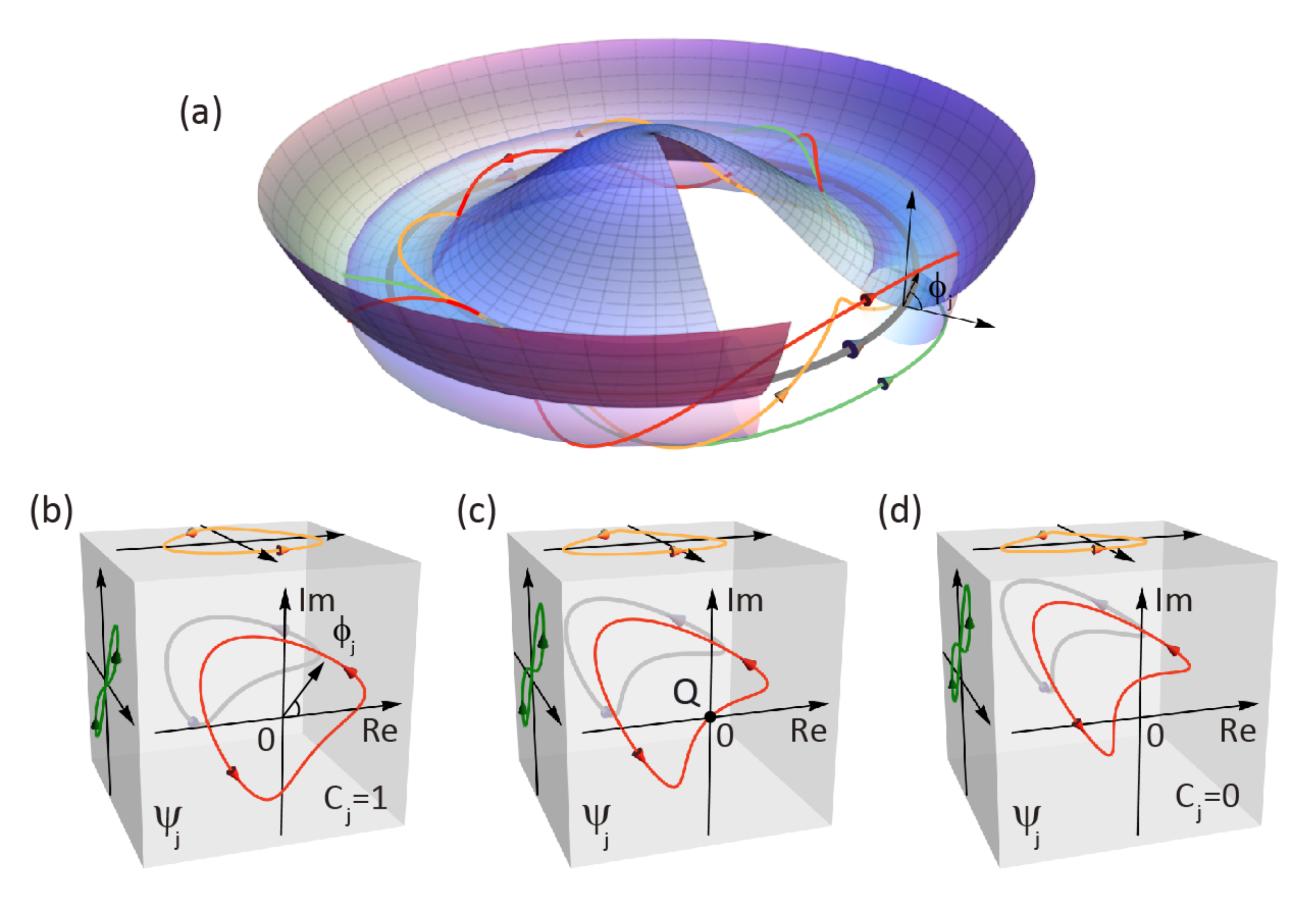}
\caption{(a) Illustration of degenerate ground states and the order-parameter trajectories. Here the "Mexican hat" represents the energy functional with respect to the order parameter. The gray bold represents the degenerate ground states, where the arrow denotes the evolution direction with respect to the group parameter. The thin colorful curves illustrate the degenerate-space trajectories of the order parameter. The translucent torus is only drawn to guide the eye to the winding angles (e.g. $\phi_j$ for component $\psi_j$). Topologically distinct trajectories of one specific order-parameter component $\psi_j$ (may be realized with different physical parameters) are shown on the front face of cubes: (b) $C_j=1$; (c) transition point (the winding is broken at the singular point $Q$); (d) $C_j=0$. }
\label{fig:phasetopo}
\end{figure}

We define the topology for each component of the ground state $\{\psi_j(\xi),\xi\in [0,2\pi)\}, j=1,\ldots,N$ [see Fig.~\ref{fig:phasetopo}(a)], as given by projecting $\Psi$ onto every $j$th axis. The $\psi_j(\xi)$ states with $\xi\in [0,2\pi)$ span a closed projection space generated by symmetry group $\mathcal{G}$. The winding number for each projected component is defined by
\begin{equation}\label{eq_TWN}
C_{j}=\frac{1}{2\pi}\oint d\xi\tilde{\psi}_{j}(\xi)^{\ast}(-i\partial_{\xi})\tilde{\psi}_{j}(\xi),\quad j=1,\ldots,N,
\end{equation}
with $\tilde{\psi}_{j}(\xi)=\psi_{j}(\xi)/|\psi_{j}(\xi)|$, and describes the winding of $\psi_j(\xi)$ around their complex-plane origin. The complete topological invariants read $\mathcal{C}=(C_{1},C_{2},\cdots,C_{N})^{T}$ for all projected components.
The change of $\mathcal{C}$ describes an emergent topological transition, e.g. three topologically distinct trajectories of $\psi_j(\xi)$ are shown in Fig.~\ref{fig:phasetopo}(b)-\ref{fig:phasetopo}(d). The winding number $C_{j}$ is unchanged unless the phase of $\psi_{j}(\xi)$ becomes ill defined when the condensate density of this component vanishes at certain point of the parameter space [the origin in Fig.~\ref{fig:phasetopo}(c)],
which mimics the gap closing in conventional topological phase transition.

We note that the topology defined above in the projection complex order parameter space, dubbed order parameter projection topology (OPPT), is different from the topology of the original symmetry group~\cite{monastyrsky2013topology,golo1993symmetry}, which is connected to the former by nonhomeomorphous projection mapping $\Psi\rightarrow\psi_j$. The projection space has richer topological structure, in which the emergent topological transition gives nontrivial physics in the collective excitations, as we study below.

{\emph{Type-II NG mode emerging from OPPT transition}}.--We show that the OPPT transition leads to the emergence of type-II NG modes. The Bogoliubov Hamiltonian at zero momentum takes generic block matric form~\cite{takahashi2015counting}
\begin{equation}\label{eq_BH}
\mathcal{H}_0=\left(\begin{array}{cc}
F & G\\
-G^{\ast} & -F^{\ast}
\end{array}\right),
\end{equation}
where $G_{ij}=\frac{\partial^{2}H_0}{\partial\psi_{i}^{\ast}\partial\psi_{j}^{\ast}}$ and $F_{ij}=\frac{\partial^{2}H_0}{\partial\psi_{i}^{\ast}\partial\psi_{j}}-\mu\delta_{ij}$, with $i,j=1,\ldots,N$. At the transition point, let the ground state be $\Psi=\Psi_Q$ at the singular point $Q$ [see Fig.~\ref{fig:phasetopo}(c)].
Since $\psi_{j}=0$ at point $Q$, we find the off-diagonal terms $F_{ij}=G_{ij}=F_{ji}=G_{ji}=0$ for $i\neq j$ from Eq.~(\ref{eq_Vint}). Further, from the GP equation we find that $F_{jj}=e^{-2i\phi_{j}}R_{j}$ and $G_{jj}=R_{j}$ with $\varphi_{j}=\arg(\psi_{j})$ and $R_{j}=\sum_{k,l\neq1}g_{jjkl}\psi_{k}\psi_{l}\big|_{\Psi\rightarrow\Psi_{Q}}$.
It'll be shown $R_{j}$ generally also vanish due to the winding breaking.

For the ground state, the energy functional stays at its minimum. Therefore
\begin{equation}\label{eq_partialphi}
-i\frac{\partial H_{0}}{\partial\phi_{j}}=2\psi_{j}^{\ast 2}\sum_{k,l\neq j}g_{jjkl}\psi_{k}\psi_{l}-\text{H.c.}=0,
\end{equation}
which implies that $\psi_{j}^{\ast 2}R_{j}$ is real. Then $R_{j}$ has phase equal to twice of that of $\psi_j$, and in general also undergoes winding breaking at point $Q$. Thus $R_{j}$ vanishes at the same critical point $Q$, giving that $F_{jj}=G_{jj}=0$. Away from the transition point, $F_{jj}$ and $G_{jj}$ are generically nonzero.
With this analysis we have that the $j$ th column and $j$ th row of $F$ and $G$ matrices vanish at the transition point. Hence $\mathcal{H}_0$ has two permanent zero solutions, $\boldsymbol{y}$ with elements of $\boldsymbol{y}_{i}=\delta_{i,j}-\delta_{i,j+N}$ and $\boldsymbol{z}=\sigma \boldsymbol{y}^{\ast}$ due to the presence of the symmetry breaking and topological transition, where only the $j$ th and $(N+j)$-th elements are nonvanishing. Here $\sigma=\sigma_{z}\otimes I_{N\times N}$ with the Pauli matrix $\sigma_z$ and the $N$-dimensional identity matrix $I_{N\times N}$. All other zero solutions of $\mathcal{H}_0$ can be normalized to have vanishing $j$ th and $(N+j)$ th elements, and are orthogonal to $\boldsymbol{y}$ and $\boldsymbol{z}$.
After orthogonalization $\boldsymbol{y}$ and $\boldsymbol{z}$ can combine into a pair of dual conjugate finite-norm zero modes, for which a type-II (also type-B~\cite{type_AB}) NG mode can be constructed in this subspace~\cite{takahashi2015counting,suplementary}.

{\emph{OPPT in the $s$-$p$ band model}}.--The above theory can be directly applied to the mixed $s$-$p$-band model, which includes three components $\psi_{j=x,y,z}$, and the transition of OPPT defined with the $\text{SO(2)}$ symmetry is responsible for the emergence of type-II NG mode at the $T$-$M$ phase boundary. The winding numbers are $C_{x,y}=1$ for both $p_x$ and $p_y$ components in subphase $T$, representing the nontrivial state. In subphase $M$, the topological invariant is, however, ill defined since the trajectories of $p_{x,y}$ components cross the origin, representing the critical state. When we approach from $T$ side to the boundary, the winding of $\psi_{j}$, well defined in the $T$ phase, breaks down right at the boundary [see Figs.~\ref{fig:PD}(c)-\ref{fig:PD}(e), where the gray markers in (e) are singular points], leading to vanishing of $F_{jj}$ and $G_{jj}$ ($j=x$ or $y$). Thus the type-II NG mode emerges at the $T$-$M$ boundary, as shown in Fig.~\ref{fig:NG MODE}.
Inside subphase $M$, the phase of $\psi_{j}^2$ keeps constant and has no singularities, so, generally, $F_{jj}, G_{jj}\neq0$ from Eq.~\eqref{eq_partialphi}, and no type-II NG mode is obtained inside subphase $M$ ($j=x$ or $y$). In this case, we have only one zero mode $\boldsymbol{y}$ with $\boldsymbol{y}_{i}=\delta_{i,j}e^{i\phi_{j}}-\delta_{i,j+N}e^{-i\phi_{j}}$ ($\phi_{j}$ takes the limit approaching $Q$)~\cite{suplementary}, and its dual nonzero mode $\boldsymbol{z}=\sigma\boldsymbol{y}^{\ast}$ satisfying $\mathcal{H}_{0}\boldsymbol{z}=2\kappa\boldsymbol{y}$ with finite $\kappa$ at point $Q$, which are the seeds of a type-I NG mode~\cite{takahashi2015counting}.

\emph{Detection}.--The emergence of the new type-II NG mode generically accompanies the softening of phonon velocity which is nonanalytic at critical point. This effect can be detected by measuring the scaling behavior without fine-tuning. We show that the dispersion $\epsilon=vk+\mathcal{O}(k^{2})$ and phonon velocity $v\propto|\gamma_{0,1}-\gamma_{0,1}^{(c)}|^{\nu}$, with the critical exponent $\nu=1/2$ and critical points $\gamma_{0,1}^{(c)}$ satisfying $(\gamma_{0}^{(c)}-\gamma_{1}^{(c)})^{2}-\tilde{\Delta}(\gamma_{0}^{(c)}+2\gamma_{1}^{(c)})/\rho=0$~\cite{suplementary}. By measuring the collective excitation spectra with proven techniques, such as the Bragg spectroscopy~\cite{ji2015softening}, the nonanalytic softening of $v$ as well as the critical exponent and critical points can be observed by fitting the data nearby phase boundary, without need of tuning exactly to the critical point. This provides an experimental verification of the emergence of type-II NG mode and the OPPT transition~\cite{sachdev2007quantum}. More details are seen in Supplemental Material~\cite{suplementary}.

\emph{Conclusion}.--We have predicted an unconventional type-II NG mode which emerges at a critical region of emergent topological transition in the mixed $s$ and $p$ band Bose superfluids. The emergence of the new type-II NG mode is beyond the symmetry-breaking mechanism, but has a new universal mechanism interpreted by the topological transition on a projective complex order-parameter space, i.e., the order-parameter projection topology on top of spontaneous symmetry breaking. The OPPT defined in the degenerate space generated by broken symmetries is conceptually different from the conventional topological phases~\cite{hasan2010colloquium}, including the topological superconductor~\cite{qi2011topological} in which the topology and symmetry-breaking order coexist but are  independent concepts
~\cite{suplementary}. 
In comparison, the unconventional type-II NG mode and its underlying topological mechanism predicted here provide a paradigm intrinsically bridging the classification of quantum matter through Landau symmetry-breaking theory and topological theory, and may open up a new direction in this field.

\emph{Acknowledgements}.--
The authors are indebted to Lin Zhang, Xiaopeng Li, Daisuke Takahashi, Wei Yi, Chao Gao and Zehan Li for helpful discussion. This work is supported by the National Postdoctoral Program for Innovative Talents of China under Grant No. BX201700156 and the
National Natural Science Foundation of China (Grant
No. 11825401, No. 11904228, No. 11804221, No. 11761161003, and No. 11921005), the National Key R$\&$D
Program of China (2016YFA0301604), the Strategic Priority Research Program of Chinese Academy of Science (Grant No. XDB28000000), the Shanghai Talent Program and the Science and Technology Commission of Shanghai Municipality
(Grants No.16DZ2260200), AFOSR Grant No.  FA9550-16-1-0006, MURI-ARO Grant No. W911NF-17-1-0323 through UC Santa Barbara, and Shanghai Municipal Science and Technology Major Project (Grant No. 2019SHZDZX01), and by the Open Project of Shenzhen Institute of Quantum Science and Engineering (Grant No. SIQSE202003).


\begin{widetext}
\section*{Supplemental Material}
In this Supplemental Material, we provide more details for the definition of interaction coefficients, operation of the $SO(2)$ rotation symmetry, numerical calculation, analytic study of low-energy excitation spectra, comparison between the order-parameter projection topology and the conventional band topology, and experimental detection scheme.

\subsection{Interaction coefficients}
Considering the symmetries of $p$-orbital functions (with odd parity for $p_\nu$-orbital in $\nu$ direction and even parity in the other two directions), the scattering processes in the band-mixed model can be classified into three classes, i.e. $\{ p_{x},p_{x}\rightarrow p_{x},p_{x};$ $ p_{y},p_{y}\rightarrow p_{y},p_{y};\quad p_{x},p_{y}\rightarrow p_{x},p_{y}\}$, $\left\{ p_{x},s\rightarrow p_{x},s;\quad p_{y},s\rightarrow p_{y},s\right\}$, and $\left\{s,s\rightarrow s,s\right\}$, which are specified with three interaction coefficients $\gamma_{j=0,1,2}=\int dzw_{a}^{2(2-j)}(z)w_{b}^{2j}(z)$ with the lowest even-parity orbital functions $w_{a}$ of the shallower layer and $w_{b}$ of the deeper layer in the z direction [see Fig. 1(a) in the main text]. Here we have already integrated the x- and y-direction degrees of freedom by employing the harmonic-oscillator approximation for the orbital functions in these directions~\cite{li2016physics}. By tuning the barrier height/width of the double-well potential, one can control the strength of $\gamma_1$ with respect to $\gamma_0$ and $\gamma_2$, while typically $\gamma_1$ should be smaller than $\gamma_0$ and $\gamma_2$. We would like to emphasize that, recently a new type of double-well potential with sub-wavelength barrier is realized with the assistance of dark states, which may be employed to realize comparable $\gamma_j$~\cite{Lacki2016Nanoscale, Wang2018Dark}. Without loss of generality, we directly set $\gamma_0=\gamma_2$ for convenience and study the phase diagram on the $\gamma_0-\gamma_1$ plane in this paper.

\subsection{Operation of the $SO(2)$ rotation symmetry on the ground state}
The generator of the $SO(2)$ rotation symmetry is $\hat{L}_{z}$, which can be represented by Pauli matrix $\sigma_y$ for the spinors defined as $(\begin{array}{cc}
\ensuremath{\langle\hat{\psi}_{x,\boldsymbol{Q}_{x}}\rangle}, & \langle\hat{\psi}_{y,\boldsymbol{Q}_{y}}\rangle\end{array})/\sqrt{N_a}=(\begin{array}{cc}
e^{i\phi_{x}}\cos\varphi, & e^{i\phi_{y}}\sin\varphi\end{array})^{T}$. The transformation by the $SO(2)$ symmetry then gives the rotation of the psuedospin, characterized by the wave-function $(\begin{array}{cc} e^{i\phi_{x}}\cos\varphi, & e^{i\phi_{y}}\sin\varphi\end{array})^{T}$, around the $y$ axis. The degenerate ground state subspace generated by the broken $SO(2)$ symmetry leads to the degenerate trajectory in the ($\phi_x,\phi_y, \varphi$)-parameter space. We can find that the pseudospin $(\begin{array}{cc} e^{i\phi_{x}}\cos\varphi, & e^{i\phi_{y}}\sin\varphi\end{array})^{T}$ of the ground state has a finite projection in the $y$ direction in sub-phase $T$ and the $SO(2)$ rotation of the pseudospin is mapped to a conical surface (i.e. all the degenerate pseudospin ground states generated by the $SO(2)$ symmetry). In contrast, the pseudospin is perpendicular to the $y$ axis in sub-phase $M$ and the rotation keeps the pseudospin in the $z-x$ plane. The $SO(2)$ rotation symmetry and the global $\text{U(1)}$ gauge symmetry clearly commute.

\subsection{Numerical calculation}
In this subsection, we give more details for the numerical calculation of mean-field ground states and low-energy excitation spectra. Replacing the field operators with the superfluid order parameters in the Hamiltonian $H$, $(\langle\hat{\psi}_{z\boldsymbol{k}}\rangle, \langle\hat{\psi}_{x\boldsymbol{k}}\rangle, \langle\hat{\psi}_{y\boldsymbol{k}}\rangle)=\sqrt{N_a}(
\delta_{\boldsymbol{k},\boldsymbol{0}}\cos\theta, \delta_{\boldsymbol{k},\boldsymbol{Q}_{x}}\sin\theta\cos\varphi e^{i\phi_{x}},  \delta_{\boldsymbol{k},\boldsymbol{Q}_{y}}\sin\theta\sin\varphi e^{i\phi_{y}})$, we derive the mean-field energy functional of the p-band superfluids
\begin{equation}\label{seq_EF}
\begin{split}
\langle H\rangle/N_{a}&=-4t\cos^{2}\theta-2(t_{\parallel}+t_{\perp})\sin^{2}\theta+\Delta\sin^{2}\theta+\frac{1}{2}\rho\gamma_{0}\cos^{4}\theta+\frac{3}{8}\rho\gamma_{0}(\sin^{4}\theta-\frac{4}{3}\sin^{4}\theta\cos^{2}\varphi\sin^{2}\varphi\sin^{2}\phi_{-})\\
&+\rho\gamma_{1}\sin^{2}\theta\cos^{2}\theta[1+\frac{1}{2}(\cos^{2}\varphi\cos2\phi_{x}+\sin^{2}\varphi\cos2\phi_{y})]\\
&=\lambda_{2}\sin^{4}\theta+\lambda_{1}\sin^{2}\theta+\lambda_{0},\\
\end{split}
\tag{S1}
\end{equation}
where $\phi_-=(\phi_y-\phi_x)$ and
\begin{equation}\label{seq_lam}
\begin{split}
&\lambda_{0}=-4t+\frac{1}{2}\rho\gamma_{0},\quad\lambda_{1}=\tilde{\Delta}-\rho\gamma_{0}+\rho\gamma_{1}[1+\frac{1}{2}(\cos^{2}\varphi\cos2\phi_{x}+\sin^{2}\varphi\cos2\phi_{y})],\\
&\lambda_{2}=\rho\{\frac{7}{8}\gamma_{0}-\frac{1}{2}\gamma_{0}\cos^{2}\varphi\sin^{2}\varphi\sin^{2}\phi_{-}-\gamma_{1}[1+\frac{1}{2}(\cos^{2}\varphi\cos2\phi_{x}+\sin^{2}\varphi\cos2\phi_{y})]\},
\end{split}
\tag{S2}
\end{equation}
with $\tilde{\Delta}=\Delta+4(t-\frac{t_{\parallel}+t_{\perp}}{2})$. In general, we search the mean-field ground state by minimizing $\langle\mathcal{H}\rangle/N_{a}$ numerically.  Benefitting from the quadratic form of $\sin^2\theta$, we can easily find the minimum of the energy functional by simply analyzing the coefficients $\lambda_{0,1,2}$.

The presence of continuous symmetry $SO(2)$ leads to the ground state is continuously degenerate. It is hard to fix all of the degenerate states in numerics completely. Fortunately, in sub-phases M and T, $\sin^2\theta=-\lambda_1/2\lambda_2^2\neq 0,1$, and then we actually can derive partially analytical forms for order parameters, which provide us a manageable way to fix the degenerate states. In this case, the minimum of energy density of $\langle H\rangle/N_{a}$ is given by $E_{min}=-\lambda_{1}^{2}/4\lambda_{2}$.  Owing to the lack of symmetry between $p_z$ and $p_x$ or $p_y$ components, there is no degeneracy for different $\sin^2\theta$. What we want to discuss is the degeneracy of the minimum in the space of by $\phi_x$, $\phi_y$ and $\varphi$. A trivial case is when $\phi_{x,y}=(2m_{x,y}+1)\pi/2$, $m_{x,y}\in\mathbb{Z}$, where $E_{min}$ is independent to $\varphi$, which corresponds to sub-phase M discussed in the main text. In the following, we will focus on the degeneracy in sub-phase T where $\sin2\phi_{-}\neq0$. To find the lowest energy for a specific $\varphi$, we require $\frac{\partial E_{min}}{\partial\phi_{x,y}}=0$, which leads to
\begin{equation}\label{seq_partialE_min}
2(\frac{\partial\lambda_{1}}{\partial\phi_{x}}\pm\frac{\partial\lambda_{1}}{\partial\phi_{y}})-\frac{\lambda_{1}}{\lambda_{2}}(\frac{\partial\lambda_{2}}{\partial\phi_{x}}\pm\frac{\partial\lambda_{2}}{\partial\phi_{y}})=0.
\tag{S3}
\end{equation}
Since
\begin{equation}\label{seq_partialphixy}
\begin{split}
&\frac{\partial\lambda_{1}}{\partial\phi_{x,y}}=\rho\gamma_{1}\cos^{2}\varphi\sin2\phi_{x,y},\quad\frac{\partial\lambda_{2}}{\partial\phi_{x,y}}=\pm\rho\gamma_{0}\cos^{2}\varphi\sin^{2}\varphi\sin\phi_{-}\cos\phi_{-}+\rho\gamma_{1}\cos^{2}\varphi\sin2\phi_{x,y},
\end{split}
\tag{S4}
\end{equation}
we further derive
\begin{equation}\label{seq_trajectory}
\begin{split}
&\cos^{2}\varphi\cos2\phi_{x}=u\sin^{2}\varphi-\frac{1}{2}(\frac{1}{u}+u),\quad\sin^{2}\varphi\cos2\phi_{y}=-u\sin^{2}\varphi-\frac{1}{2}(\frac{1}{u}-u),\\
&\cos^{2}\varphi\sin2\phi_{x}+\sin^{2}\varphi\sin2\phi_{y}=0,
\end{split}
\tag{S5}
\end{equation}
where $u=\frac{2\gamma_{1}(\gamma_{1}-\frac{3}{4}\gamma_{0}-\tilde{\Delta}/\rho)}{\gamma_{1}^{2}+\gamma_{0}(\frac{3}{2}\gamma_{1}-\gamma_{0}+\tilde{\Delta}/\rho)}$. Substituting Eq.~(\ref{seq_trajectory}) to the energy functional (\ref{seq_EF}), we can find $E_{min}$ doesn't depend on $\varphi$. It means $E_{min}$ given by any $\varphi$ satisfying Eq~(\ref{seq_trajectory}) is degenerate. The trajectory of degenerate states in Fig. 1 in the main text can be plotted with the Eq.~(\ref{seq_trajectory}) following the procedure: 1) for a specific $\varphi$, calculate $\phi_{x,y}$ with Eq.~(\ref{seq_trajectory}); 2) calculate $\lambda_{0,1,2}$ with Eq.~(\ref{seq_lam}); 3) fix $\theta$ with the minimum condition $\sin^2\theta=-\lambda_1/2\lambda_2^2$. For example, when $\varphi=\pi/4$, we have $\cos(2\phi_x)=\cos(2\phi_y)=-1/u$ and $\sin(2\phi_x)=-\sin(2\phi_y)$, which fixes the possible values of $\phi_{x}$ and $\phi_y$. Further, we can fix $\theta$ through $\sin^2\theta=-\lambda_1/2\lambda_2^2$ and finally access the degenerate mean-field ground states.
Fig. 1(b)-(e) in the main text are plotted in this way.

The Bogoliubov spectra shown in Fig. 2 in the main text is calculated by diagonalizing the Bogoliubov Hamiltonian $\mathcal{H}_b$ (elaborated below) using paraunitary transformations, i.e. numerically solving the eigenvalue equation $\sigma\mathcal{H}_b\boldsymbol{w}=\epsilon\boldsymbol{w}$. Here $\sigma=\sigma_{z}\otimes I_{3\times3}$, $\boldsymbol{w}$ is the eigenvector, and $\epsilon$ is the eigenvalues. Since the system possesses the translation symmetry, the spectra can be calculated separately for different Bloch quasi-momentum $\boldsymbol{k}$. The dispersion coefficients, which are shown in Fig. 2a in the main text, are approximately given by the slopes of the spectra at $\boldsymbol{k}=0$ point. The lattice size is chosen as $80\times 80$ in the numerical calculation.

\subsection{Analytical study of low-energy excitation spectra}
We provide the details on the perturbation analysis of low-energy excitation spectra in this subsection. In the theory of Takahashi and Nitta~\cite{takahashi2015counting}, zero solutions of Bogoliubov Hamiltonian with zero (finite) norm are the seed of type-I (type-II) Nambu-Goldstone (NG) mode. After considering the momentum-dependent terms, which act as perturbation in the low-momentum limit, these solutions will give rise to the spectra of the NG modes. Although the norms of zero solutions $\boldsymbol{y}$ and $\boldsymbol{z}$ arising at topological transition are all zero, i.e. $\boldsymbol{y}^\dagger\sigma\boldsymbol{y}=\boldsymbol{z}^\dagger\sigma\boldsymbol{z}=0$, finite-norm
zero solutions $(\boldsymbol{y}\pm\boldsymbol{z})/2$ can be constructed and a type-II NG mode emerges. We will illustrate this point with the following analysis on the example model.
\subsubsection{Bogoliubov Hamiltonian}
The Hamiltonian of the band-mixed model is given by $H=H_0+H_T$, where $H_T$ includes the hopping terms and $H_0$ includes the on-site terms. The minimum of the three bands of $H_T$ are at different quasi-momentum points. For convenience, we employ the transformation
\begin{equation}\label{seq:newframe}
\left(\begin{array}{c}
\hat{\psi}_{0\boldsymbol{j}}\\
\hat{\psi}_{1\boldsymbol{j}}\\
\hat{\psi}_{2\boldsymbol{j}}
\end{array}\right)=\frac{1}{\sqrt{2}}\left(\begin{array}{c}
\sqrt{2}\hat{\psi}_{z\boldsymbol{j}}\\
\left(-1\right)^{j_{x}}\hat{\psi}_{x\boldsymbol{j}}-i\left(-1\right)^{j_{y}}\hat{\psi}_{y\boldsymbol{j}}\\
\left(-1\right)^{j_{x}}\hat{\psi}_{x\boldsymbol{j}}+i\left(-1\right)^{j_{y}}\hat{\psi}_{y\boldsymbol{j}}
\end{array}\right),
\tag{S6}
\end{equation}
to shift the minimums of the energy bands to $\boldsymbol{k}=0$~\cite{li2011effective}. The forms of $H_T$ and $H_0$ then are cast into
\begin{equation}\label{seq_T}
H_T=\sum_{\lambda,\lambda^{'}=1,2}\sum_{\langle\boldsymbol{i},\boldsymbol{j}\rangle}\hat{\psi}_{\lambda\boldsymbol{i}}^{\dagger}T_{\lambda\boldsymbol{i},\lambda^{'}\boldsymbol{j}}\hat{\psi}_{\lambda^{'}\boldsymbol{j}},
\tag{S7}
\end{equation}
and
\begin{equation}\label{seq_H0}
\begin{split}
H_{0}=&\frac{1}{2}\sum_{\boldsymbol{j}}\{\gamma_{0}\hat{n}_{\boldsymbol{j}}^{(0)}(\hat{n}_{\boldsymbol{j}}^{(0)}-1)+\gamma_{1}[2\hat{n}_{\boldsymbol{j}}^{\left(0\right)}\hat{n}_{\boldsymbol{j}}^{\left(12\right)}+(\hat{\psi}_{0\boldsymbol{j}}^{\dagger2}\hat{\psi}_{1\boldsymbol{j}}\hat{\psi}_{2\boldsymbol{j}}+H.c.)]+\frac{3}{4}\gamma_{2}[\hat{n}_{\boldsymbol{j}}^{(12)}(\hat{n}_{\boldsymbol{j}}^{(12)}-\frac{2}{3})-\frac{1}{3}\hat{L}_{z,\boldsymbol{j}}^{2}]\\
&+2\Delta\hat{n}_{\boldsymbol{j}}^{(12)}-2\mu\hat{n}_{\boldsymbol{j}}\},
\end{split}
\tag{S8}
\end{equation}
where $\langle\dots\rangle$ restricts the summation to run over the nearest neighbour sites, $\hat{n}_{\boldsymbol{j}}^{(0)}=\hat{n}_{\boldsymbol{j}}^{(z)}$, $\hat{n}_{\boldsymbol{j}}=\sum_{\lambda=0,1,2}\hat{\psi}_{\lambda\boldsymbol{j}}^{\dagger}\hat{\psi}_{\lambda\boldsymbol{j}}$, $\hat{n}_{\boldsymbol{j}}^{\left(12\right)}=\sum_{\lambda=1,2}\hat{\psi}_{\lambda\boldsymbol{j}}^{\dagger}\hat{\psi}_{\lambda\boldsymbol{j}}$, $\hat{L}_{z,\boldsymbol{j}}=\left(-1\right)^{j_{x}+j_{y}}(\hat{\psi}_{1\boldsymbol{j}}^{\dagger}\hat{\psi}_{1\boldsymbol{j}}-\hat{\psi}_{2\boldsymbol{j}}^{\dagger}\hat{\psi}_{2\boldsymbol{j}})$, and
\begin{equation}\label{seq_TM}
T_{\boldsymbol{j},\boldsymbol{j}\pm\boldsymbol{1}_{x}}=T_{x}=\left(\begin{array}{ccc}
-t & 0 & 0\\
0 & -\frac{t_{\parallel}+t_{\perp}}{2} & -\frac{t_{\parallel}-t_{\perp}}{2}\\
0 & -\frac{t_{\parallel}-t_{\perp}}{2} & -\frac{t_{\parallel}+t_{\perp}}{2}
\end{array}\right),\quad T_{\boldsymbol{j},\boldsymbol{j}\pm\boldsymbol{1}_{y}}=T_{y}=\left(\begin{array}{ccc}
-t & 0 & 0\\
0 & -\frac{t_{\parallel}+t_{\perp}}{2} & \frac{t_{\parallel}-t_{\perp}}{2}\\
0 & \frac{t_{\parallel}-t_{\perp}}{2} & -\frac{t_{\parallel}+t_{\perp}}{2}
\end{array}\right).
\tag{S9}
\end{equation}

In the quasimomentum space, we have $H_T=2\sum_{\boldsymbol{k}}\hat{\Psi}_{\boldsymbol{k}}^{\dagger}(T_{x}\cos k_{x}+T_{y}\cos k_{y})\hat{\Psi}_{\boldsymbol{k}}$, where $\hat{\Psi}_{\boldsymbol{k}}=(\begin{array}{ccc}
\hat{\psi}_{1\boldsymbol{k}} & \hat{\psi}_{2\boldsymbol{k}} & \hat{\psi}_{3\boldsymbol{k}}\end{array})^{T}=N^{-1/2}\sum_{\boldsymbol{j}}(\begin{array}{ccc}
\hat{\psi}_{0\boldsymbol{j}} & \hat{\psi}_{1\boldsymbol{j}} & \hat{\psi}_{2\boldsymbol{j}}\end{array})^{T}e^{-i\boldsymbol{k}\cdot\boldsymbol{j}}$. In general, $t_{\parallel}$ is larger than $t_{\perp}$, and then the minimums of the three bands all local at $\boldsymbol{k}=0$. The ground-state superfluid order parameter in the new basis is given by $\langle\hat{\psi}_{\nu\boldsymbol{k}}\rangle=\delta_{\boldsymbol{k},\boldsymbol{0}}\psi_{\nu},\nu=0,1,2$, where
\begin{equation}\label{seq_GS}
\left(\begin{array}{c}
\psi_{0}\\
\psi_{1}\\
\psi_{2}
\end{array}\right)=N_{a}^{1/2}\left(\begin{array}{ccc}
1 & 0 & 0\\
0 & 1 & -i\\
0 & 1 & i
\end{array}\right)\left(\begin{array}{c}
\cos\theta\\
\sin\theta\cos\varphi e^{i\phi_{x}}\\
\sin\theta\sin\varphi e^{i\phi_{y}}
\end{array}\right).
\tag{S10}
\end{equation}

The Bogoliubov Hamiltonian can be derived by expanding the field operator $\hat{\Psi}_{\boldsymbol{j}}=(\begin{array}{ccc}
\hat{\psi}_{0\boldsymbol{j}} & \hat{\psi}_{1\boldsymbol{j}} & \hat{\psi}_{2\boldsymbol{j}}\end{array})^{T}$ into its mean field and fluctuation $\hat{\Psi}_{\boldsymbol{j}}=\langle\hat{\Psi}_{\boldsymbol{j}}\rangle+\delta\hat{\Psi}_{\boldsymbol{j}}$~\cite{pethick2008bose}. Under the quasi-momentum Nambu basis $(\delta\hat{\Psi}_{\boldsymbol{k}}^{\ast},\delta\hat{\Psi}_{-\boldsymbol{k}})^{T}$ with $\delta\hat{\Psi}_{\boldsymbol{k}}=N^{-1/2}\sum_{\boldsymbol{j}}\delta\hat{\Psi}_{\boldsymbol{j}}e^{-i\boldsymbol{k}\cdot\boldsymbol{j}}=(\begin{array}{ccc}
\delta\hat{\psi}_{0\boldsymbol{k}} & \delta\hat{\psi}_{1\boldsymbol{k}} & \delta\hat{\psi}_{2\boldsymbol{k}}\end{array})^{T}$,
the Bogoliubov Hamiltonian is given by $H_{b}=(\delta\hat{\Psi}_{\boldsymbol{k}}^{\dagger},\delta\hat{\Psi}_{-\boldsymbol{k}}^{T})(\mathcal{H}_{0}\delta_{\boldsymbol{k},0}+\mathcal{H}_{T})(\delta\hat{\Psi}_{\boldsymbol{k}}^{\ast},\delta\hat{\Psi}_{-\boldsymbol{k}})^{T}$, where $\mathcal{H}_0$ is given by Eq. (6) in the main text with
\begin{equation}\label{seq_F}
F=\left(\begin{array}{ccc}
2\gamma_{0}n_{0}+\gamma_{1}n_{12}-\mu & \gamma_{1}\left(\psi_{0}\psi_{1}^{\ast}+\psi_{0}^{\ast}\psi_{2}\right) & \gamma_{1}\left(\psi_{0}\psi_{2}^{\ast}+\psi_{0}^{\ast}\psi_{1}\right)\\
\gamma_{1}\left(\psi_{0}^{\ast}\psi_{1}+\psi_{0}\psi_{2}^{\ast}\right) & \gamma_{0}n_{12}+\gamma_{1}n_{0}+\tilde{\Delta}-\mu & \gamma_{0}\psi_{1}\psi_{2}^{\ast}\\
\gamma_{1}\left(\psi_{0}^{\ast}\psi_{2}+\psi_{0}\psi_{1}^{\ast}\right) & \gamma_{0}\psi_{1}^{\ast}\psi_{2} & \gamma_{0}n_{12}+\gamma_{1}n_{0}+\tilde{\Delta}-\mu
\end{array}\right),
\tag{S11}
\end{equation}
and
\begin{equation}\label{seq_G}
G=\left(\begin{array}{ccc}
\gamma_{0}\psi_{0}^{2}+\gamma_{1}\psi_{1}\psi_{2} & \gamma_{1}\psi_{0}\psi_{1} & \gamma_{1}\psi_{0}\psi_{2}\\
\gamma_{1}\psi_{0}\psi_{1} & \frac{1}{2}\gamma_{0}\psi_{1}^{2} & \gamma_{0}\psi_{1}\psi_{2}+\frac{1}{2}\gamma_{1}\psi_{0}^{2}\\
\gamma_{1}\psi_{0}\psi_{2} & \gamma_{0}\psi_{1}\psi_{2}+\frac{1}{2}\gamma_{1}\psi_{0}^{2} & \frac{1}{2}\gamma_{0}\psi_{2}^{2}
\end{array}\right).
\tag{S12}
\end{equation}
Here $n_{0}=\left|\psi_{0}\right|^{2}$ and $n_{12}=\left|\psi_{1}\right|^{2}+\left|\psi_{2}\right|^{2}$. Note that we have already set $\gamma_2=\gamma_0$ in the above equations. The chemical potential $\mu$ is given by
\begin{equation}
\mu=\left(\begin{array}{ccc}
\psi_{0}^{\ast} & \psi_{1}^{\ast} & \psi_{2}^{\ast}\end{array}\right)\left[\left(\begin{array}{ccc}
\gamma_{0}n_{0}+\gamma_{1}n_{12} & \frac{1}{2}\gamma_{1}\psi_{0}^{\ast}\psi_{2} & \frac{1}{2}\gamma_{1}\psi_{0}^{\ast}\psi_{1}\\
\frac{1}{2}\gamma_{1}\psi_{0}\psi_{2}^{\ast} & \frac{1}{2}\gamma_{0}n_{12}+\gamma_{1}n_{0}+\tilde{\Delta} & \frac{1}{2}\gamma_{0}\psi_{1}\psi_{2}^{\ast}\\
\frac{1}{2}\gamma_{1}\psi_{0}^{\ast}\psi_{1} & \frac{1}{2}\gamma_{0}\psi_{1}^{\ast}\psi_{2} & \frac{1}{2}\gamma_{0}n_{12}+\gamma_{1}n_{0}+\tilde{\Delta}
\end{array}\right)\right]\left(\begin{array}{c}
\psi_{0}\\
\psi_{1}\\
\psi_{2}
\end{array}\right).
\tag{S13}
\end{equation}
$\mathcal{H}_T$ corresponds to the fluctuation expansion of $H_T$ and takes the form
\begin{equation}\label{eq:BogoH_T}
\mathcal{H}_{T}=2\left(\begin{array}{cc}
T_{x}\cos(k_{x}a)+T_{y}\cos(k_{y}a) & 0\\
0 & -[T_{x}\cos(k_{x}a)+T_{y}\cos(k_{y}a)]
\end{array}\right).
\tag{S14}
\end{equation}
In the long-wave limit $\boldsymbol{k}\rightarrow\boldsymbol{0}$, we have the approximation $\mathcal{H}_{b}\approx K_{0}\delta_{\boldsymbol{k},\boldsymbol{0}}+K_{2}k^2$, where
\begin{equation}\label{eq:K0}
K_{0}=\left(\begin{array}{cccc}
-4t & 0 & 0 & 0\\
0 & \left[-2\left(t_{\parallel}+t_{\perp}\right)\right]I_{2} & 0 & 0\\
0 & 0 & 4t & 0\\
0 & 0 & 0 & \left[2\left(t_{\parallel}+t_{\perp}\right)\right]I_{2}
\end{array}\right)+\mathcal{H}_{0},
\tag{S15}
\end{equation}
and
\begin{equation}\label{eq:K2x}
K_{2}=\left(\begin{array}{cccc}
2t & 0 & 0 & 0\\
0 & \left(t_{\parallel}+t_{\perp}\right)I_{2}+\left(t_{\parallel}-t_{\perp}\right)\sigma_{x}\cos(2\phi_{k}) & 0 & 0\\
0 & 0 & -2t & 0\\
0 & 0 & 0 & -\left(t_{\parallel}+t_{\perp}\right)I_{2}-\left(t_{\parallel}-t_{\perp}\right)\sigma_{x}\cos(2\phi_{k})
\end{array}\right),
\tag{S16}
\end{equation}
with $\phi_k=\arctan(k_y/k_x)$. Considering the $k^2$ term as the perturbation, this form is convenient for the perturbation analysis of NG modes.

The low-energy excitation spectra, as the eigen-spectra of Bogoliubov Hamiltonian, can be calculated with a perturbation theory by setting the momentum-independent part $K_0$ and momentum-dependent part $K_{2}k^2$ respectively as the zero-order Hamiltonian and perturbation Hamiltonian~\cite{takahashi2015counting}. The basic idea of the perturbation theory of Bogoliubov spectra are similar to the conventional perturbation theory, i.e. 1) solving the eigenvalue problem of the zero-order Hamiltonian; 2) expanding the eigenvalues and eigenvectors of the Bogoliubov Hamiltonian into different perturbation orders; 3) deriving the perturbative eigenvalues and eigenvectors by matching different perturbation orders in the two sides of the full eigenvalue equation. This procedure has been elaborated in details in Ref.~\cite{takahashi2015counting}. In the following, we will focus on the analysis in sub-phases $M$ and $T$, which analytically confirms the emergence of type-II NG modes at the boundary between sub-pases $M$ and $T$.

\subsubsection{Zero solutions of Bogoliubov Hamiltonian}
The properties of the zero solutions of momentum-independent part of Bogoliubov Hamiltonian $K_0$ are crucial for the dispersion of NG modes. Since the low-energy Hamiltonian of our model possesses two continuous symmetries, the global $\text{U(1)}$ gauge symmetry and $\text{SO(2)}$ rotation symmetry, the number of NG modes in general is smaller than 2 according to the conventional symmetry-based argument~\cite{nielsen1976count,watanabe2011number,watanabe2012unified,hidaka2013counting,watanabe2014effective, leutwyler1994nonrelativistic,schafer2001kaon}. The two NG modes are constructed with two pairs of zero-norm modes $\boldsymbol{y}_j$ and $\boldsymbol{z}_j$, $j=1,2$~\cite{takahashi2015counting}, which satisfies the relations
\begin{equation}\label{seq_yz}
\begin{split}
&K_{0}\boldsymbol{y}_{j}=0,\quad K_{0}\boldsymbol{z}_{j}=2\kappa_{j}\boldsymbol{y}_{j},\\
&\boldsymbol{y}_{j}^{\dagger}\sigma\boldsymbol{y}_{j}=\boldsymbol{z}_{j}^{\dagger}\sigma\boldsymbol{z}_{j}=\boldsymbol{y}_{j}^{\dagger}\sigma\boldsymbol{y}_{3-j}=\boldsymbol{z}_{j}^{\dagger}\sigma\boldsymbol{z}_{3-j}=\boldsymbol{y}_{j}^{\dagger}\sigma\boldsymbol{z}_{3-j}=0,\quad\boldsymbol{y}_{j}^{\dagger}\sigma\boldsymbol{z}_{j}=\boldsymbol{z}_{j}^{\dagger}\sigma\boldsymbol{y}_{j}=2,
\end{split}
\tag{S17}
\end{equation}
where $\sigma=\sigma_{z}\otimes I_{3\times 3}$ with Pauli matrix $\sigma_z$ and 3-by-3 identity matrix $I_{3\times 3}$. Here $\boldsymbol{y}_j$ are the zero solutions of $K_0$ and $\boldsymbol{z}_j$ are their dual modes.

Without loss of generality, we will analyze the NG modes above the degenerate ground states with $\varphi=\pi/4$, i.e. the superfluid order parameters take the forms $\Psi^{(M)}=(\cos\theta, \frac{i}{\sqrt{2}}\sin\theta e^{i\pi/4},\frac{i}{\sqrt{2}}\sin\theta e^{-i\pi/4})^T$ in the sub-phase $M$ and $\Psi^{(T)}=(\cos\theta, \frac{1}{\sqrt{2}}\sin\theta\left(\cos\phi-\sin\phi\right)e^{-i\pi/4},\frac{1}{\sqrt{2}}\sin\theta\left(\cos\phi+\sin\phi\right)e^{i\pi/4})^T$ in the sub-phase $T$. It is worthy to emphasize that, the dispersion properties of NG modes are independent of the degenerate state taken in the calculation. Then the symmetry-generated zero modes are given by $\boldsymbol{y}_{1}^{(M,T)}=(\Psi^{(M,T)},-\Psi^{(M,T)*})^{T},\quad\boldsymbol{y}_{2}^{(M,T)}=[I_{2\times 2}\otimes(0\oplus\sigma_{z})]\boldsymbol{y}_{1}^{(M,T)}$.
Modes $\boldsymbol{y}_{1}^{(M,T)}$ and $\boldsymbol{y}_{2}^{(M,T)}$ are the zero modes corresponding to the $\text{U(1)}$ and $\text{SO(2)}$ symmetries. Approaching the $T-M$ phase boundary, $\phi\rightarrow\pm\pi/2$, and then $\Psi^{(T)}\rightarrow\Psi^{(M)}$ and $\boldsymbol{y}_{1,2}^{(T)}\rightarrow\boldsymbol{y}_{1,2}^{(M)}$.

After lengthy calculation based on Eq.~(\ref{seq_yz}), we can derive the dual modes of $\boldsymbol{y}_{1,2}^{(M,T)}$ are given by $\boldsymbol{z}_{1,2}^{(M,T)}=(\chi_{1,2}^{(M,T)},\chi_{1,2}^{(M,T)\ast})^{T}$, where
\begin{equation}
\begin{split}
\chi_{1}^{(M)}=&(\begin{array}{ccc}
\lambda_{10}^{(M)}\cos\theta, & \frac{i}{\sqrt{2}}\lambda_{11}^{(M)}\sin\theta e^{i\pi/4}, & \frac{i}{\sqrt{2}}\lambda_{12}^{(M)}\sin\theta e^{-i\pi/4}\end{array}),\\
\chi_{2}^{(M)}=&\frac{1}{\sin\theta}(\begin{array}{ccc}
0, & \frac{i}{\sqrt{2}}e^{i\pi/4}, & -\frac{i}{\sqrt{2}}e^{-i\pi/4}\end{array}),\\
\chi_{1}^{(T)}=&(\lambda_{10}^{(T)}\begin{array}{ccc}
\cos\theta, & \frac{1}{\sqrt{2}}\lambda_{11}^{(T)}\sin\theta\left(\cos\phi-\sin\phi\right)e^{-i\pi/4}, & \frac{1}{\sqrt{2}}\lambda_{12}^{(T)}\sin\theta\left(\cos\phi+\sin\phi\right)e^{i\pi/4}\end{array}),\\
\chi_{2}^{(T)}=&(\begin{array}{ccc}
\lambda_{20}^{(T)}\cos\theta, & \frac{1}{\sqrt{2}}\lambda_{21}^{(T)}\sin\theta\left(\cos\phi-\sin\phi\right)e^{-i\pi/4}, & \frac{1}{\sqrt{2}}\lambda_{22}^{(T)}\sin\theta\left(\cos\phi+\sin\phi\right)e^{i\pi/4}\end{array}),
\end{split}
\tag{S18}
\end{equation}
with
\begin{equation}
\begin{split}
&\lambda_{10}^{(M)}=\frac{\frac{3\gamma_{0}}{2}-\gamma_{1}\sin^{2}\theta}{\cos^{2}\theta(\frac{3\gamma_{0}}{2}-\gamma_{1}\sin^{2}\theta)+\sin^{2}\theta(2\gamma_{0}-\gamma_{1}\cos^{2}\theta)},\\
&\lambda_{11}^{(M)}=\lambda_{12}^{(M)}=\frac{2\gamma_{0}-\gamma_{1}\cos^{2}\theta}{\cos^{2}\theta(\frac{3\gamma_{0}}{2}-\gamma_{1}\sin^{2}\theta)+\sin^{2}\theta(2\gamma_{0}-\gamma_{1}\cos^{2}\theta)},\\
&\lambda_{10}^{(T)}=\frac{1}{\cos^{2}\theta+\frac{1}{2}r_{1}\sin^{2}\theta+\frac{1}{2}s_{1}\sin^{2}\theta\sin2\phi},\quad\lambda_{11,12}^{(T)}=\frac{r_{1}\mp s_{1}}{2\cos^{2}\theta+r_{1}\sin^{2}\theta+s_{1}\sin^{2}\theta\sin2\phi},\\
&\lambda_{20}^{(T)}=-\frac{2}{\sin^{2}\theta\left(r_{2}\sin2\phi+s_{2}\right)},\quad\lambda_{21,22}^{(T)}=\frac{r_{2}\pm s_{2}}{2\sin^{2}\theta\left(r_{2}\sin2\phi+s_{2}\right)}.
\end{split}
\tag{S19}
\end{equation}
Here $r_{1,2}=\frac{B_{1,2}A_{1,2}^{'}-C_{1,2}B_{1,2}^{'}}{A_{1,2}B_{1,2}^{'}-B_{1,2}A_{1,2}^{'}}$ and $s_{1,2}=-\frac{A_{1,2}C_{1,2}^{'}-C_{1,2}A_{1,2}^{'}}{A_{1,2}B_{1,2}^{'}-B_{1,2}A_{1,2}^{'}}$, where
\begin{equation}
\begin{split}
&A_{1}=\sin^{2}\theta\cos2\phi\left[\frac{3}{4}\gamma_{0}-\gamma_{1}\left(1+\frac{1}{2}\cos2\phi\right)\right],\quad B_{1}=\sin2\phi\left[\left(\frac{3}{4}\gamma_{0}-\gamma_{1}\right)\sin^{2}\theta\cos2\phi+\frac{1}{2}\gamma_{1}\cos^{2}\theta\right],\\
&C_{1}=\cos^{2}\theta\left[\gamma_{1}\left(1+2\cos2\phi\right)-2\gamma_{0}\cos2\phi\right],\\
&A_{1}^{'}=\frac{1}{2}\gamma_{0}\sin^{2}\theta\cos2\phi\sin2\phi,\quad B_{1}^{'}=\frac{1}{2}\gamma_{0}\sin^{2}\theta\cos2\phi+\gamma_{1}\cos^{2}\theta,\quad C_{1}^{'}=2\gamma_{1}\cos^{2}\theta\sin2\phi,\\
&A_{2}=\gamma_{1}\sin^{2}\theta\left(1+\frac{1}{2}\cos2\phi\right),\quad B_{2}=\gamma_{1}\sin^{2}\theta\sin2\phi,\quad C_{2}=2\gamma_{0}\cos^{2}\theta,\\
&A_{2}^{'}=\frac{3}{4}\gamma_{0}\sin^{2}\theta\cos2\phi,\quad B_{2}^{'}=\sin2\phi\left[\frac{3}{4}\gamma_{0}\sin^{2}\theta\cos2\phi+\frac{1}{2}\gamma_{1}\cos^{2}\theta\right],\quad C_{2}^{'}=\gamma_{1}\cos^{2}\theta\left(1+2\cos2\phi\right).
\end{split}
\tag{S20}
\end{equation}
Correspondingly, the $\kappa$ functions are given by
\begin{equation}\label{eq:kappas}
\begin{split}
&\kappa_{1}^{(M)}=\frac{3\gamma_{0}}{4}\lambda_{11}^{(M)}\sin^{2}\theta+\frac{\gamma_{1}}{2}\lambda_{10}^{(M)}\cos^{2}\theta,\\
&\kappa_{2}^{(M)}=-\frac{1}{4}\gamma_{0}+\frac{1}{2}\gamma_{1}\cot^{2}\theta,\\
&\kappa_{1}^{(T)}=\lambda_{10}^{(T)}\gamma_{0}\cos^{2}\theta+\frac{1}{2}\lambda_{10}^{(T)}\gamma_{1}\sin^{2}\theta\left[r_{1}\left(1+\frac{1}{2}\cos2\phi\right)+s_{1}\sin2\phi\right],\\
&\kappa_{2}^{(T)}=\frac{1}{2}\lambda_{20}^{(T)}\gamma_{1}\cos^{2}\theta\frac{2\cos\phi+\left(\cos\phi-\sin\phi\right)}{\cos\phi-\sin\phi}+\frac{1}{4}\lambda_{21}^{(T)}\left[\gamma_{0}\left(1-\sin2\phi\right)\sin^{2}\theta-\gamma_{1}\cos^{2}\theta\frac{\cos\phi+\sin\phi}{\cos\phi-\sin\phi}\right]\\
&\quad\quad\quad+\frac{1}{2}\lambda_{22}^{(T)}\frac{\cos\phi+\sin\phi}{\cos\phi-\sin\phi}\left(\gamma_{0}\cos2\phi\sin^{2}\theta+\frac{\gamma_{1}}{2}\cos^{2}\theta\right).
\end{split}
\tag{S21}
\end{equation}
Approaching the $T-M$ phase boundary, $\boldsymbol{z}_{1,2}^{(T)}\rightarrow\boldsymbol{z}_{1,2}^{(M)}$. At the $T-M$ phase boundary, $\kappa_{2}^{(M)}=\kappa_{2}^{(T)}=0$, which means $\boldsymbol{z}_2$ becomes a zero mode of $\mathcal{H}_b$ and the type-I NG mode corresponding to the $\text{SO(2)}$ symmetry softens to a type-II NG mode.

\subsubsection{Perturbation expansion of Bogoliubov eigenvalue equation}
Now let us go foward to the perturbation analysis of the diagonalization of Bogoliubov
Hamiltonian.
Assuming $H_b\boldsymbol{\xi}=\epsilon\boldsymbol{\xi}$, and taking the perturbation expansions
\begin{equation}\label{eq:expxi}
\begin{split}
\boldsymbol{\xi}=&\boldsymbol{\xi}_{0}+k\boldsymbol{\xi}_{1}+k^{2}\boldsymbol{\xi}_{2}+k^{3}\xi_{3}\cdots,\\
\epsilon=&v_{0}+kv_{1}+k^{2}v_{2}+k^{3}v_{3}\cdots,
\end{split}
\tag{S22}
\end{equation}
we derive the lowest-three-order perturbation equations
\begin{equation}\label{eq:pereq}
\begin{split}
K_{0}\boldsymbol{\xi}_{1}=&v_{1}\boldsymbol{\xi}_{0},\\
K_{0}\boldsymbol{\xi}_{2}+K_{2}\boldsymbol{\xi}_{0}=&v_{1}\boldsymbol{\xi}_{1}+v_{2}\boldsymbol{\xi}_{0},\\
K_{0}\boldsymbol{\xi}_{3}+K_{2}\boldsymbol{\xi}_{1}=&v_{1}\boldsymbol{\xi}_{2}+v_{2}\boldsymbol{\xi}_{1}+v_{3}\boldsymbol{\xi}_{0}.
\end{split}
\tag{S23}
\end{equation}
Here $v_{1}$ is the phonon velocity.

Assuming $\boldsymbol{w}_{1}$ is the last non-negative-energy eigenvector of $K_{0}$ besides $\boldsymbol{y}_{1}$ and $\boldsymbol{y}_{2}$, which in general corresponds to a finite eigenvalue $\varepsilon_1$, we yield
\begin{equation}
\boldsymbol{\xi}_{0}=c_{1}^{\left(0\right)}\boldsymbol{y}_{1}+c_{2}^{\left(0\right)}\boldsymbol{y}_{2},
\quad\boldsymbol{\xi}_{j}=d_{1}^{\left(j\right)}\boldsymbol{z}_{1}+d_{2}^{\left(j\right)}\boldsymbol{z}_{2}+\alpha_{1}^{\left(j\right)}\boldsymbol{w}_{1}+\beta_{1}^{\left(j\right)}\tau\boldsymbol{w}_{1}^{\ast},\quad j=1,2,
\tag{S24}
\end{equation}
The first order equations become
\begin{equation}
\left(2\kappa_{1}d_{1}^{(1)}-c_{1}^{(0)}v_{1}\right)\boldsymbol{y}_{1}+\left(2\kappa_{2}d_{2}^{(1)}-c_{2}^{\left(0\right)}v_{1}\right)\boldsymbol{y}_{2}+\varepsilon_{1}(\alpha_{1}^{(1)}\boldsymbol{w}_{1}+\beta_{1}^{(1)}\tau\boldsymbol{w}_{1}^{\ast})=0,
\tag{S25}
\end{equation}
Then we obtain
\begin{equation}\label{S26}
2\kappa_{1}d_{1}^{\left(1\right)}-c_{1}^{\left(0\right)}v_{1}=0,\quad2\kappa_{2}d_{2}^{\left(1\right)}-c_{2}^{\left(0\right)}v_{1}=0,\quad\alpha_{1}^{\left(1\right)}=\beta_{1}^{\left(1\right)}=0.
\tag{S26}
\end{equation}
It means $\boldsymbol{\xi}_{1}$ include only the $\boldsymbol{z}_{i}$
component:
\begin{equation}
\boldsymbol{\xi}_{j}=v_{1}\frac{c_{1}^{(0)}}{2\kappa_{1}}\boldsymbol{z}_{1}+v_{1}\frac{c_{2}^{(0)}}{2\kappa_{2}}\boldsymbol{z}_{2}.
\tag{S27}
\end{equation}

In order to determine $v_{1}$, $c_{1}^{(0)}$
and $c_{2}^{(0)}$, we mutiply the second-order perturbation equations
from left with $\boldsymbol{y}_{1}^{\dagger}\sigma$ and $\boldsymbol{y}_{2}^{\dagger}\sigma$,
then derive
\begin{equation}
\boldsymbol{y}_{i}^{\dagger}\sigma K_{2}\boldsymbol{\xi}_{0}=v_{1}\boldsymbol{y}_{i}^{\dagger}\sigma\boldsymbol{\xi}_{1}+v_{2}\boldsymbol{y}_{i}^{\dagger}\sigma\boldsymbol{\xi}_{0},\quad i=1,2.
\tag{S28}
\end{equation}
On the other hand, we can derive
\begin{equation}
\boldsymbol{y}_{1}^{\dagger}\sigma K_{2}\boldsymbol{\xi}_{0}=c_{1}^{\left(0\right)}\boldsymbol{y}_{1}^{\dagger}\sigma K_{2}\boldsymbol{y}_{1}+c_{2}^{\left(0\right)}\boldsymbol{y}_{1}^{\dagger}\sigma K_{2}\boldsymbol{y}_{2}=c_{1}^{\left(0\right)}\left[2t\cos^{2}\theta+\left(t_{\parallel}+t_{\perp}\right)\sin^{2}\theta\right],
\tag{S29}
\end{equation}
and
\begin{equation}
\boldsymbol{y}_{2}^{\dagger}\sigma K_{2}\boldsymbol{\xi}_{0}=c_{1}^{\left(0\right)}\boldsymbol{y}_{2}^{\dagger}\sigma K_{2}\boldsymbol{y}_{1}+c_{2}^{\left(0\right)}\boldsymbol{y}_{2}^{\dagger}\sigma K_{2}\boldsymbol{y}_{2}=c_{2}^{\left(0\right)}\left(t_{\parallel}+t_{\perp}\right)\sin^{2}\theta.
\tag{S30}
\end{equation}
Then we yield
\begin{equation}
c_{1}^{\left(0\right)}\left[2t\cos^{2}\theta+\left(t_{\parallel}+t_{\perp}\right)\sin^{2}\theta\right]=2v_{1}d_{1}^{(1)},
\tag{S31}
\end{equation}
and
\begin{equation}
c_{2}^{\left(0\right)}\left(t_{\parallel}+t_{\perp}\right)\sin^{2}\theta=2v_{1}d_{2}^{(1)}.
\tag{S32}
\end{equation}

Combining with Eqs.~(S26), when $c_{1}^{\left(0\right)}=1$ and $c_{2}^{\left(0\right)}=0$, we have
\begin{equation}
v_{1}=v_{U(1)}=\sqrt{2\kappa_{1}\left[t\cos^{2}\theta+\left(t_{\parallel}+t_{\perp}\right)\sin^{2}\theta/2\right]},\quad d_{1}^{(1)}=\sqrt{\frac{t\cos^{2}\theta+\left(t_{\parallel}+t_{\perp}\right)\sin^{2}\theta/2}{2\kappa_{1}}}.
\tag{S33}
\end{equation}
Therefore, the eigenvector and eigenvalue of the excitation mode at around the NG mode associated with the $\text{U(1)}$ symmetry are respectively given by
\begin{equation}
\boldsymbol{\xi}_{U(1)}=\boldsymbol{y}_{1}+\sqrt{\frac{t\cos^{2}\theta+\left(t_{\parallel}+t_{\perp}\right)\sin^{2}\theta/2}{2\kappa_{1}}}k\boldsymbol{z}_{1}+\mathcal{O}\left(k^{3}\right),
\tag{S34}
\end{equation}
and
\begin{equation}
\epsilon_{U(1)}=\sqrt{2\kappa_{1}\left[t\cos^{2}\theta+\left(t_{\parallel}+t_{\perp}\right)\sin^{2}\theta/2\right]}k+\mathcal{O}\left(k^{2}\right).
\tag{S35}
\end{equation}
We will show below the second-order term of $k$ in the excitation spectrum is vanishing.

When $c_{1}^{\left(0\right)}=0$ and $c_{2}^{\left(0\right)}=1$, we
derive
\begin{equation}
v_{1}=v_{SO(2)}=\sqrt{\kappa_{2}\left(t_{\parallel}+t_{\perp}\right)\sin^{2}\theta},\quad d_{2}^{(1)}=\frac{1}{2}\sqrt{\frac{\left(t_{\parallel}+t_{\perp}\right)\sin^{2}\theta}{\kappa_{2}}},
\tag{S36}
\end{equation}
Then the eigenvector and eigenvalue of the excitation mode at around the NG mode associated with the $\text{SO(2)}$ symmetry are respectively given by
\begin{equation}\label{eq:perter_wf}
\boldsymbol{\xi}_{SO(2)}=\boldsymbol{y}_{2}+\frac{1}{2}\sqrt{\frac{\left(t_{\parallel}+t_{\perp}\right)\sin^{2}\theta}{\kappa_{2}}}k\boldsymbol{z}_{1}+\mathcal{O}\left(k^{2}\right),
\tag{S37}
\end{equation}
and
\begin{equation}\label{eq:perter_en}
\epsilon_{SO(2)}=\sqrt{\kappa_{2}\left(t_{\parallel}+t_{\perp}\right)\sin^{2}\theta}k+\mathcal{O}\left(k^{2}\right).
\tag{S38}
\end{equation}

As we mentioned above, the second-order dispersion of these type-I NG modes actually doesn't exist. For $\text{U(1)}$ and $\text{SO(2)}$ modes, multiplying the second-order and third-order perturbation equations from left with $\boldsymbol{z}_{j=1,2}^{\dagger}\sigma$ and $\boldsymbol{y}_{j=1,2}^{\dagger}\sigma$, respectively, we derive $\boldsymbol{z}_{j}^{\dagger}\sigma K_{0}\boldsymbol{\xi}_{2}+\boldsymbol{z}_{j}^{\dagger}\sigma K_{2}\boldsymbol{\xi}_{0}=2v_{2}$ and $d_{j}^{(1)}\boldsymbol{y}_{j}^{\dagger}\sigma K_{2}\boldsymbol{z}_{j}=v_{1}\boldsymbol{y}_{j}^{\dagger}\sigma\boldsymbol{\xi}_{2}+2d_{j}^{(1)}v_{2}$.
Thus, we have $v_{2}=\text{Re}(\boldsymbol{y}_{j}^{\dagger}\sigma K_{2}\boldsymbol{z}_{j})/2$. Since $\boldsymbol{y}_{j}$, $\boldsymbol{z}_{j}$ and $K_{2}$ take the forms $\boldsymbol{y}_{j}=(\tilde{\boldsymbol{y}}_{j}, -\tilde{\boldsymbol{y}}^{\ast}_{j})^{T}$, $\boldsymbol{z}_{j}=(\tilde{\boldsymbol{z}}_{j}, \tilde{\boldsymbol{z}}^{\ast}_{j})^{T}$, and $K_{2}=\sigma_{z}\otimes \tilde{K}_{2}$, we derive $\boldsymbol{y}_{j}^{\dagger}\sigma K_{2}\boldsymbol{z}_{j}=\tilde{\boldsymbol{y}}_{j}^{\dagger}\tilde{K}_{2}\tilde{\boldsymbol{z}}_{j}-\tilde{\boldsymbol{z}}_{j}^{\dagger}\tilde{K}_{2}\tilde{\boldsymbol{y}}_{j}=2i\text{Im}(\tilde{\boldsymbol{y}}_{j}^{\dagger}\tilde{K}_{2}\tilde{\boldsymbol{z}}_{j})$
and then $v_{2}=0$. Therefore, the second-order dispersion of type-I NG modes in general vanishes.

Numerical calculation shows that $\kappa_1$ always keeps finite at around the boundary between sub-phases $M$ and $T$, which implies the NG mode associated with the global gauge $\text{U(1)}$ symmetry always has linear dispersion. In contrast, $\kappa_{2}$ tends to 0 near the $T-M$ boundary, which corresponds to the vanishing of the first-order dispersion coefficient of the NG mode associated with the $\text{SO(2)}$ symmetry. When $\kappa_{2}$ vanishes, we need to consider the second-order perturbation equation in the second row of Eq.~(\ref{eq:pereq}). From $v_{1}=0$, we derive $\xi_{1}=0$. In addition, because $\sigma K_{0}\boldsymbol{z}_{2}=0$, $\boldsymbol{z}_{2}$ is also one zero solution of $K_{0}$ and we need to set $\boldsymbol{\xi}_{0}=x_{2}$, where
\begin{equation}
\boldsymbol{x}_{2}=\frac{\boldsymbol{y}_{2}+\boldsymbol{z}_{2}}{2}=\left(\begin{array}{cccccc}
0 & \frac{i}{\sqrt{2}}e^{i\pi/4} & -\frac{i}{\sqrt{2}}e^{-i\pi/4} & 0 & 0 & 0\end{array}\right)^{T}.
\tag{S39}
\end{equation}
Besides, we have
\begin{equation}
K_{0}\boldsymbol{\xi}_{2}+K_{2}\boldsymbol{\xi}_{0}=v_{2}\boldsymbol{\xi}_{0}.
\tag{S40}
\end{equation}
Multiply $\boldsymbol{x}_{2}^{\dagger}\sigma$ to the above equation,
noting that $\boldsymbol{y}_{1}^{\dagger}\sigma K_{0}\boldsymbol{\xi}_{2}=0$,
we obtain
\begin{equation}
\boldsymbol{x}_{2}^{\dagger}\sigma K_{2}\boldsymbol{x}_{2}=v_{2}\boldsymbol{x}_{2}^{\dagger}\sigma\boldsymbol{x}_{2}.
\tag{S41}
\end{equation}
On the other hand, because $\boldsymbol{x}_{2}^{\dagger}\sigma K_{2}\boldsymbol{x}_{2}=\frac{t_{\parallel}+t_{\perp}}{2}$ and $x_{2}^{\dagger}\sigma x_{2}=1$, the second-order coefficient of dispersion is given by
\begin{equation}
v_{2}=\frac{t_{\parallel}+t_{\perp}}{2}.
\tag{S42}
\end{equation}
Then the dispersion of the type-II NG mode at the boundary between sub-phases T and M is given by
\begin{equation}
\epsilon_{SO(2)}=\frac{t_{\parallel}+t_{\perp}}{2}k^{2}+\mathcal{O}(k^{3}).
\tag{S43}
\end{equation}
Here, since $\boldsymbol{\xi}_{0}=\boldsymbol{x}_{1}$, $\boldsymbol{\xi}_{1}=0$, $v_{1}=0$, the third-order perturbation equation gives rise to $K_0\xi_{3}=v_{3}\boldsymbol{x}_{1}$. Multiplying this equation with $\boldsymbol{\boldsymbol{x}}^{\dagger}\sigma$ from left, we derive $v_{3}=0$, which means the third-order dispersion of type-II NG mode is vanishing.

\begin{table*}
\begin{tabular}{|c|c|c|c|c|c|}
\hline
 & base space & objective space & topological invariant & critical point & spectra\tabularnewline
\hline
\hline
band topology & Brillouin zone & Bloch states & Zak phase & gap closing & edge mode\tabularnewline
\hline
OPPT & group parameters & projection of degenerate space & winding number & modulus vanishing & type-II NG mode\tabularnewline
\hline
\end{tabular}
\caption{Comparison between the OPPT defined with a first-order symmetry group and the conventional one-dimensional fermion band topology.}
\label{stb_cp}
\end{table*}

\subsection{Comparison between order-parameter projection topology and fermion band topology}
In the main text, the theory of order-parameter projection topology (OPPT) is proposed to explain the emergent type-II NG mode. Unlike the conventional topological phases~\cite{hasan2010colloquium,qi2011topological,shen2012topological,hasan2011three,chiu2016classification,bansil2016colloquium, wen2017colloquium}, e.g. the topological insulators and topological superconductors, which are defined on the bulk (quasiparticle) band which is gapped and not degenerate, the OPPT is defined on the projection of degenerate space generated by the spontaneously broken symmetries. We provide more details on the comparison between these two concepts in this subsection.

The OPPT is defined upon the degenerate space generated by spontaneous symmetry breaking. In a symmetry-spontaneous-breaking phase, a degenerate space is formed by the degenerate ground states~\cite{monastyrsky2013topology}. The topological invariants are defined on the projection of degenerate space onto a specific basis, i.e. the manifolds formed by the mapping between the symmetry group and components of order parameter. For the case of first-order symmetry group (labeled by one group parameter), the projection manifolds are close curves on the complex plane, and the topological invariant are the winding numbers, as illustrated in Fig.~3 in the main text. It is worth noting that, OPPT is not the topology of degenerate space, which is equivalent to the topology of symmetry group, instead it is the topology defined for the projection of degenerate space on a specific basis. In presence of topological transition, the winding numbers need to be ill-defined and then the curves must cross the complex-plane origin. Particularly, combining with the Bogoliubov theory, it is shown that type-II NG modes generally emerge at the critical point of OPPT transition.

In contrast, for the conventional topological phases, the topological invariants are defined on the many-body ground states formed by the Bloch wave functions of gapped energy bands (excluding some special gapless nodal phases). Specifically, the topological space is the fiber bundle constructed by the mapping from the Brillouin zone to the Bloch states up to a gauge invariance. The energy gap above the occupied band closes at the critical point of a topological phase transition. As a consequence, topological edge modes emerge at the boundary of topological matter. Even for topological superfluid/superconductor~\cite{qi2011topological}, the pairing order parameter breaking the gauge symmetry only acts as a coupling field (like spin-orbit coupling), and the degenerate degrees of freedom are irrelevant.

Therefore, the concept of OPPT proposed in the current work is completely different from the conventional topological phases. It represents a paradigm which intrinsically bridges the Landau-symmetry breaking and topological theories, and also hosts nontrivial new physics. The comparison between the OPPT and conventional topological phase is illustrated in table~\ref{stb_cp}.

\begin{figure*}[tbp]
\includegraphics[width=11cm]{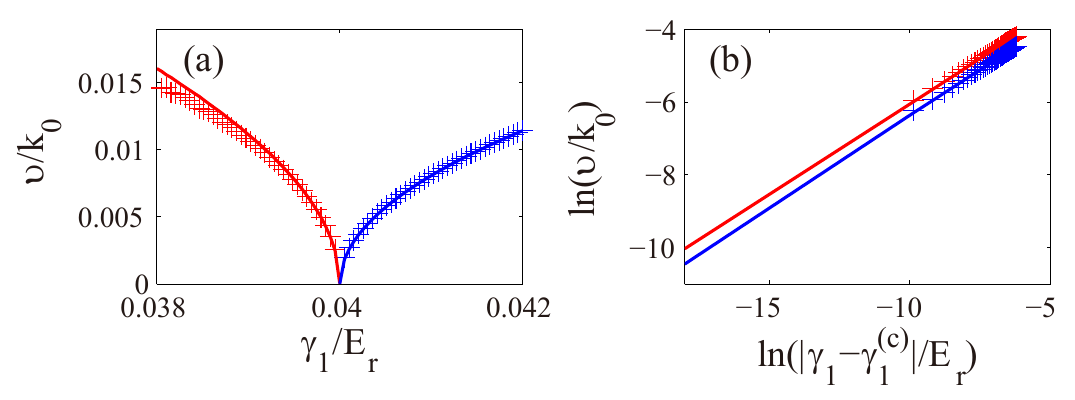}
\caption{(a) The critical behavior of phonon velocity $v$ at around the sub-phase boundary $T-M$. The crosses and solid curves represent the results calculated with the direct diagonalization of Bogoliubov Hamiltonian and the partially analytical forms in Eq.~(\ref{eq:kappas}), respectively. The colors specify different sub-phases (i.e. $T$ and $M$). From the double-log plot in (b), we can find the critical exponent is indeed $1/2$, which is consistent with the analytical results in Eqs.~(\ref{eq:kappa_2_M}) and (\ref{eq:kappa_2_T}). Here $\rho\gamma_{0}/E_{r}=0.1$, which corresponds to $\rho\gamma_{1}^{(c)}/E_{r}=0.04$. The other parameters are set to be the same with those in Fig. 1 in the main text.}
\label{fig:critical}
\end{figure*}

\subsection{Experimental detection}
We provide more details on the detection scheme for the emergence of type-II NG mode and OPPT transition based on the non-analytical vanishing of phonon velocity in this subsection. As shown above, the dispersion of NG mode associated with the $\text{SO(2)}$ symmetry is given by $\epsilon=v k+\mathcal{O}\left(k^{2}\right)$ (the subscript of $\text{SO(2)}$ is quitted for convenience), with the phonon velocity $v=\sqrt{\kappa_{2}\left(t_{\parallel}+t_{\perp}\right)\sin^{2}\theta}$. In fact, in sub-phase $M$, since $\phi_{x,y}=\pm \pi/2$, we can simplify the form of $\kappa_{2}$ in sub-phase $M$, $\kappa_{2}^{(M)}$, in Eq.~(\ref{eq:kappas}) into the form
\begin{equation}\label{eq:kappa_2_M}
\kappa_{2}^{(M)}=\frac{\left(\gamma_{0}-\gamma_{1}\right)^{2}-\tilde{\Delta}\left(\gamma_{0}+2\gamma_{1}\right)/\rho}{2(\gamma_{1}-2\gamma_{0}+2\tilde{\Delta}/\rho)}.
\tag{S44}
\end{equation}
On the other hand, approaching the critical point, we derive
\begin{equation}\label{eq:kappa_2_T}
\kappa_{2}^{(T)}\approx\frac{\rho\gamma_{0}(\gamma_{0}^{2}-3\gamma_{1}^{2})[\gamma_{0}^{2}-3\gamma_{1}^{2}+(\tilde{\Delta}/\rho)(2\gamma_{1}-\gamma_{0})][(\gamma_{0}-\gamma_{1})^{2}-(\tilde{\Delta}/\rho)(2\gamma_{1}+\gamma_{0})]}{2(3\gamma_{0}^{2}-\gamma_{1}^{2})(\gamma_{1}^{2}-\gamma_{0}^{2}+\gamma_{0}\gamma_{1}+\gamma_{0}\tilde{\Delta}/\rho)^{2}}.
\tag{S45}
\end{equation}
At the critical point, we find $(\gamma_{0}^{(c)}-\gamma_{1}^{(c)})^{2}-\tilde{\Delta}(\gamma_{0}^{(c)}+2\gamma_{1}^{(c)})/\rho=0$, but the other terms in $\kappa_{2}^{(M,T)}$ are finite, which means $\kappa_{2}^{(M,T)}$ simultaneously softens to zero. Therefore, we have $v\propto|\gamma_{0,1}-\gamma_{0,1}^{(c)}|^{\nu}$ with the critical exponent $\nu=1/2$ at around the critical points $\gamma_{0,1}^{(c)}$, as shown in Fig.~\ref{fig:critical}. It means $v$ continuously softens to zero, as a consequence of the emergence of type-II NG mode, at the critical point of OPPT with non-analytical behavior. In experiment, the collective excitation spectra can be measured with proven techniques like the Bragg spectroscopy~\cite{ji2015softening}. The phonon velocity $v$ is given by the slope of dispersion at around the gapless point. Then the nonanalytic softening of $v$ as well as the critical exponents and critical points can be detected by fitting the data at around the transition points with the above analytical form of $v$. This detection scheme will give us the signature for the emergence of type-II NG mode, as well as the transition of OPPT~\cite{sachdev2007quantum}, without the fine tuning of parameters.

\end{widetext}
\bibliographystyle{apsrev4-1}
\bibliography{DSTopo}

\begin{thebibliography}{48}
\bibitem [{ord()}]{orderparameter}%
  \BibitemOpen
  \href@noop {} {\ }\bibinfo {note} {Here the global phase responsible for the
  breaking of the $U(1)$ symmetry is suppressed.}\BibitemShut {Stop}%
\bibitem [{mod()}]{modelconnection}%
  \BibitemOpen
  \href@noop {} {\ }\bibinfo {note} {The multiband lattice model in Eqs. (1)
  and (2) can be mapped to a homogeneous multicomponent model in the
  thermodynamic limit, which captures the low-energy physics of bosons
  condensed at around the band bottoms.}\BibitemShut {Stop}%
\end{thebibliography}

\end{document}